\documentclass[12pt]{article}
\usepackage{graphicx}
\usepackage{jheppub2}
\usepackage{amsmath,amssymb,amscd,braket,amsfonts,mathrsfs}
\usepackage{color}
\usepackage[table]{xcolor}

\usepackage{bigfoot}
\usepackage{slashed}
\usepackage{amsthm}
\usepackage{verbatim}
\usepackage{mathtools}
\usepackage{bbold}
\usepackage[linesnumbered,ruled]{algorithm2e}
\usepackage{placeins}

\title{
Turbulent gravitational wakes
}

\author[a]{Giulio Taiocchi,}
\emailAdd{giulio.taiocchi@ucdconnect.ie}

\author[a]{Christiana Pantelidou}
\affiliation[a]{School of Mathematics and Statistics, University College Dublin, Belfield, Dublin 4, Ireland}
\emailAdd{christiana.pantelidou@ucd.ie}

\author[b,c]{and Miguel Zilhão}
\affiliation[b]{Department of Physics, University of Aveiro, 3810-193 Aveiro, Portugal}
\affiliation[c]{Center for Research and Development in Mathematics and Applications (CIDMA), \\
University of Aveiro, 3810-193 Aveiro, Portugal}
\emailAdd{mzilhao@ua.pt}

\abstract{We construct turbulent gravitational wakes in asymptotically AdS$_4$ spacetimes by solving the full non-linear Einstein equations in the characteristic formulation. Specifically, we show that a localized boundary deformation acting on a boosted black brane generates a wake whose boundary dynamics exhibit the formation of coherent vortical structures that are advected downstream by the background flow. Analysing a wake propagating along the $x$ direction, we show that it displays the characteristic self-similar scaling of classical turbulent wakes. We then identify the geometric imprint of the wake on the horizon area density, the traceless horizon extrinsic curvature, and bulk metric perturbations, finding coherent structures that extend from the asymptotic boundary to the horizon. These results provide the first study of turbulent gravitational wakes in four-dimensional asymptotically AdS spacetimes.}

\begin{document}
\maketitle

\section{Introduction}\label{sec:intro}

Turbulence is one of the most ubiquitous and striking manifestations of nonlinear dynamics in nature. It appears across an enormous range of scales and physical systems, from atmospheric and oceanic flows to astrophysical plasmas, stellar interiors, and the quark-gluon plasma produced in heavy-ion collisions. Despite centuries of study, turbulence remains among the most challenging unsolved problems in theoretical physics. Its defining feature is the emergence of universal behaviour from strongly nonlinear dynamics, most notably the transfer of energy across scales through turbulent cascades and the appearance of characteristic power-law spectra; the most widely celebrated results on the universality of turbulent cascades are captured by Kolmogorov's theory of 1941 \cite{K41a, K41b}. While turbulence was originally studied in the context of non-relativistic fluids, the modern understanding of the phenomenon extends far beyond conventional hydrodynamics. In recent years, developments in the AdS/CFT correspondence~\cite{Maldacena:1997re} have suggested that turbulent behaviour may also arise in gravitational systems, opening the possibility that spacetime itself can exhibit fluid-like turbulent dynamics~\cite{Carrasco:2012nf, Green:2013zba}.

Early studies of turbulence in the holographic context focused on decaying turbulence in the hydrodynamic regime, constructing the turbulent AdS geometry via the fluid/gravity gradient expansion \cite{Carrasco:2012nf, Green:2013zba}. Particular focus was placed on conformal viscous and inviscid fluids\footnote{This is to be contrasted to the majority of the fluid turbulence literature that focuses on non-relativistic fluids.} that live either on flat or spherical spaces, monitoring, among other quantities, the behaviour of the energy power spectrum, velocity structure functions and also the decay rate of quasinormal modes (QNMs). It was seen that, in the turbulent regime, some modes grow exponentially fast and interrupt the expected exponential decay of the mode used to drive the system out of equilibrium through the initial conditions. 

A first step toward going beyond the hydrodynamic limit was taken in \cite{Adams:2013vsa}, which demonstrated that black holes in asymptotically AdS$_4$ spacetimes can exhibit turbulent behaviour at intermediate timescales when evolved from unstable initial conditions using the full non-linear Einstein equations in the characteristic formulation; at late times they relax to equilibrium. The study focused on the power spectrum of the rescaled traceless horizon curvature, and even though evidence of an inverse cascade was seen (transfer of energy from shorter to longer scales) compatible with the phenomenology of turbulence in 2+1 dimensions, Kolmogorov's scaling exponent was not seen due to the lack of driving. This work was also the first to ask which geometric quantities carry an imprint of turbulence.

More recently, the idea of gravitational turbulence was explored in the context of the large-$D$ limit of general relativity in asymptotically AdS$_D$ spacetimes~\cite{ Emparan:2015hwa,Emparan:2015gva,Bhattacharyya:2015dva, Emparan:2016sjk}. At large $D$, there is a separation of scales between the black hole size, $r_0$, and the region occupied by a nontrivial gravitational potential, $r_0/D$. A separation of scales signals an effective theory, which can be constructed by analytically solving the integrals for radial evolution. What remains from Einstein's equations is a set of constraint equations in $D-1$ dimensions that resemble fluid-dynamics equations. Crucially, however, even though they are perturbative in $1/D$, they are exact in gradients. Therefore, far from a mere technical simplification, the $1/D$ expansion allows us to directly connect black holes with the turbulent behaviour of a class of fluid-like equations that are exact in gradients. This should be contrasted with the usual treatment of hydrodynamics in a dynamical setting, where one typically truncates at a finite order and treats the resulting system of equations as exact, a procedure which fundamentally changes the theory. As a consequence of this change one may discover physically undesirable qualities such as instabilities and acausal behaviour both for relativistic~\cite{PhysRevD.31.725} and non-relativistic~\cite{Poovuttikul:2019ckt} theories. Furthermore, one must of course also verify post-hoc that the solution remained a good approximation within the framework of a perturbative gradient expansion. 

Within the large-$D$ context, \cite{Rozali:2017bll} studied decaying turbulence in asymptotically AdS spacetime restricting the dynamics in 2+1 and 3+1 dimensions and focusing on the energy power spectrum. In both cases turbulence was observed:
in 2+1 dimensions the cascade was towards longer scales (inverse cascade), while in 3+1 towards shorter scales (direct cascade). Expanding in this direction, \cite{Andrade:2019rpn} studied  turbulence in 2+1 dimensions driven by a homogeneous and isotropic forcing function that injected vorticity in the system in the large-$D$ limit. This study demonstrated that forced black hole horizons exhibit statistically steady turbulent spacetime dynamics consistent with Kolmogorov's theory\footnote{Note that the Kraichnan-Kolmogorov in two spatial dimensions and the Kolmogorov scaling in three spatial dimensions, strictly speaking, are valid for non-relativistic incompressible fluids. Generalisations of these to relativistic systems have not been explored much in the literature, with the exception of a handful of attempts. In particular, \cite{Westernacher-Schneider:2015gfa, Westernacher-Schneider:2017snn} derived scaling relations for two-point functions of the stress-tensor for relativistic fluid flows at low Mach numbers, which were subsequently tested numerically. However, at low Mach number, or equivalently flows with velocities below the speed of sound, the relativistic corrections are still small.}, predicting an inverse energy cascade from shorter to longer scales  with a specific $k^{-5/3}$ scaling exponent, this was the first time that Kolmogorov scaling was found for black hole spacetimes (outside the hydrodynamic limit). It was also demonstrated that tidal deformations of the horizon induce turbulent dynamics in the form of wakes. When set in motion relative to the horizon, a deformation develops a turbulent spacetime wake, indicating that turbulent spacetime dynamics may play a role in binary mergers and other strong-field phenomena. Neither of these results could have been obtained without driving the system.  

To add further relevance to the scenario of turbulent tidal wakes, the calculation of \cite{Andrade:2019rpn} may be taken in the same spirit as studies of near-extremal Kerr black holes and their near-horizon regions using conformal field theory (CFT). There, when a massive body falls into the near horizon region it appears as a source term in the CFT.  Thus, the sources used for driving the system may be viewed as the gravitational deformation due to such a massive body, and the appearance of a turbulent wake in this context may affect the plunge dynamics and associated waveforms for black hole or black hole-neutron star mergers.

More recently, \cite{Waeber:2021xba, Oz:2024smd} extended the results from driven turbulence in the large-$D$ limit to $D=4$ spacetime dimensions. Specifically, they studied homogeneous and isotropic driven gravitational turbulence in AdS$_4$ in the presence of random external sources in the metric using stochastic methods.  When the forcing is done through the space-space component of the metric they observe an inverse energy cascade characterised by $k^{-5/3}$ (incompressible modes dominate). When it is done through the time-time component, they find that the cascade is characterised by  $k^{-1.9}$ (compressible modes dominate), which is in agreement with results for compressible fluids \cite{Scott2007Nonrobustness}. In both cases, they associate the cascade with a scaling behaviour in the horizon area power spectrum. In addition, \cite{Du:2025vgc} studied $D=4$ homogeneous and isotropic turbulence sourced by a scalar field. In this case, \cite{Du:2025vgc} found a compressible energy dominated flow, with corresponding scaling power law $k^{-1.79}$. These results are consistent with the picture established in \cite{Westernacher-Schneider:2015gfa, Westernacher-Schneider:2017snn}, which showed that for two-dimensional compressible relativistic fluids the $k^{-5/3}$ scaling is not robust: the spectral exponent depends on the degree of compressibility of the forcing and on whether the direct cascade is well-resolved, with more compressible forcing and better resolution of the direct cascade driving the exponent toward $k^{-2}$.

In this paper, we extend the results of \cite{Andrade:2019rpn} for turbulent wakes from the large-$D$ limit to AdS$_4$. This is achieved with the Julia Einstein Characteristic Code (Jecco), an infrastructure written in the Julia programming language to solve Einstein's equations in the characteristic formulation~\cite{Bea:2022mfb}.
One of the benefits of using a characteristic code is that in this formalism the non-linear coupled partial differential equations reduce to a nested set of ordinary differential equations, which are much simpler to solve~\cite{Winicour:2012znc,Chesler:2013lia}.
The original Jecco code was developed and tested in~\cite{Bea:2022mfb}
and it has already been used in different settings, in particular in the study of the dynamics of phase transitions~\cite{Bea:2021zol,Bea:2021ieq,Bea:2021zsu,Bea:2024bls}.
The code is openly available on GitHub and was modified substantially to tackle the case of interest for the present paper (though its basic infrastructure remained mostly untouched).

Our main result is the construction and analysis of turbulent gravitational wakes in asymptotically AdS$_4$ spacetimes. By introducing a localized deformation on the boundary and evolving the full non-linear Einstein equations, we show that a boosted black brane develops a wake characterised by vortex formation and detachment, symmetry breaking, and the emergence of chaotic dynamics downstream of the source. Using boundary observables, we demonstrate that the resulting flow exhibits the characteristic signatures of turbulent wakes familiar from fluid dynamics, including self-similar profiles, a wake width that grows \textit{approximately} as $y_c\propto x^{1/2}$, and a maximum velocity deficit that decays \textit{approximately}  as $\bar{u}_x^m\propto x^{-1/2}$.
We then identify the corresponding geometric imprint in the bulk spacetime by studying the horizon area density, the traceless part of the horizon extrinsic curvature, and radial profiles of metric components. These quantities reveal that the wake is encoded in localized deformations of the horizon and in coherent bulk structures that extend from the boundary all the way to the near-horizon region. Taken together, our results provide the first detailed study of turbulent gravitational wakes in four-dimensional asymptotically AdS spacetimes and establish a direct connection between classical wake phenomenology and fully non-linear gravitational dynamics.

The remainder of this paper is organised as follows. In Section \ref{sec:setup} we describe the gravitational setup, including the evolution equations and a discussion on residual gauge fixing. Section \ref{sec:numerics} outlines the evolution algorithm that we implement and discusses the boundary driving and initial data used in the simulations for turbulent wakes. In Section \ref{sec:results} we present our results and in Section \ref{sec:disc} we conclude with a discussion of implications and open directions.

\section{Setup}
\label{sec:setup}

We consider the four-dimensional Einstein-Hilbert action
\begin{align}\label{eq:bulk_action}
S&=\int d^4 x \sqrt{-g} \left( R+6 \right),
\end{align}
where we have set $16\pi G=c=1$  and fixed the cosmological constant to be $\Lambda=-3$ for convenience. The variation of this action gives rise to the following field equations 
\begin{align}\label{eq:eom}
&R_{\mu\nu}+3g_{\mu\nu}=0\,,
\end{align}
which admit a large family of solutions that asymptotically approach four-dimensional AdS space with unit radius. In what follows we are interested in studying a subclass of these configurations with planar horizon that exhibit turbulent behaviour in 2 spatial dimensions, by simulating non-linear dynamics in 3+1 dimensions. Following~\cite{Andrade:2019rpn, Oz:2024smd}, we will induce turbulence by turning on a  deformation on the boundary of AdS$_4$, focusing on turbulent wakes. We will work at finite volume assuming periodic boundary conditions in the spatial directions.

\subsection{Evolution equations}
Following \cite{Chesler:2010bi,Chesler:2013lia,Attems:2017zam,Bea:2022mfb, Balasubramanian:2013yqa}, we will use the characteristic formulation of General Relativity to solve Einstein's equations in asymptotically AdS spaces. In this approach the hyperbolic partial differential equations reduce to ordinary differential equations along the characteristics. 

We will consider a coordinate system based on a family of ingoing null hypersurfaces emanating from a timelike section $\mathcal{T}$, where the null coordinate $v$ labels these  hypersurfaces and the coordinates $x,y$ label the remaining coordinates on $\mathcal{T}$. The coordinate $r$ is an affine parameter labelling the points along the null rays. In this coordinate system, the spacetime metric can be written in the following form
\begin{align}\label{eq:ansatz}
&ds^2=-A dv^2+2 dv\, dr+2F_i\, dv \,dx^i+\Sigma^2 h_{ij} dx^i \,dx^j\,,
\end{align}
where $i,j=1,2$ and 
\begin{equation}
h=\begin{pmatrix}
e^{-B} \cosh{\theta}&\sinh{\theta}\\
\sinh{\theta} &e^{B} \cosh{\theta}
\end{pmatrix}\,.
\end{equation}
Here $A,B,\Sigma, F_i, \theta$ are functions of  $(v,r,x,y)$. Note that we denote by $v$ the (ingoing) null bulk coordinate; at the boundary, $v$ becomes the usual time coordinate.  Given this ansatz, Einstein's equations reduce to the following set of nested ODEs
\begin{equation}\label{eq:eqns}
\begin{aligned}
&\Sigma''+\frac{\Sigma}{4}\left(\cosh^2{\theta} B'^2+\theta'^2\right)=0\,,\\
& F_i''=\mathcal{F}(\Sigma,B,  \theta, F_i')\,,\\
&d_+\Sigma'=\mathcal{F}(\Sigma,B,  \theta, F_1,F_2,d_+\Sigma)\,,\\
&d_+B'=\mathcal{F}(\Sigma,B,  \theta, F_1,F_2,d_+\Sigma,d_+\theta,d_+B)\,,\\
&d_+\theta'=\mathcal{F}(\Sigma,B,  \theta, F_1,F_2,d_+\Sigma,d_+\theta,d_+B)\,,\\
&A''=\mathcal{F}(\Sigma,B,  \theta, F_1,F_2,d_+\Sigma,d_+\theta,d_+B)\,,\\
&d_+F_i'=\mathcal{F}(\Sigma,B,  \theta, F_1,F_2,d_+\Sigma,d_+\theta,d_+B,d_+F_1,d_+F_2,A)\,,\\
&d_+d_+\Sigma=\mathcal{F}(\Sigma,B,  \theta, F_1,F_2,d_+\Sigma,d_+\theta,d_+B,d_+F_1,d_+F_2,A)\,,
\end{aligned}
\end{equation}
where $\mathcal{F}$ are complicated functions of the specified variables too lengthy to include here. In this order, these equations can all be solved sequentially as radial ODEs of the form
\begin{align*}
\left [A_f
(v,r,x,y)\partial_r^2 +B_f(v,r,x,y)\partial_r+C_f(v,r,x,y) \right]f(v,r,x,y)=-S_f(v,r,x,y)
\end{align*}
where the coefficients $A_f, B_f, C_f, S_f$ are fully determined. In deriving these equations, we find it convenient to define
\begin{align}\label{eq:notation}
f' &=\partial_r  \,f\\
d_+\,f & =\partial_v \,f+\frac{1}{2}\,A \partial_r \,f \label{eq:d+f}\\
d_i\,f & =\partial_{i} \,f-F_i \,\partial_r \,f\,.
\end{align}
The above correspond, respectively, to the directional derivatives along infalling radial null geodesics, outgoing radial null geodesics and along the directions orthogonal to both radial geodesics. We also find it convenient for the numerical implementation to define $z=1/r$. In terms of this coordinate, the AdS$_4$ boundary is located at $z=0$.

\subsection{Residual gauge fixing}
\label{subsec:gauge}

Looking at the ansatz, it is simple to see the invariance under the following transformation
\begin{align*}
r&\to r+\xi(v,x,y)\,,\\
A&\to A+2 \partial_v \xi(v,x,y)\,,\\
F_i&\to F_i-\partial_{i}\xi(v,x,y)\,,
\end{align*}
for any function $\xi$.
A convenient choice is to treat $\xi$ as another evolved variable and fix its equation of motion by requiring the position of the apparent horizon to be fixed at constant $z$ coordinate, $z=z_h$. This is guaranteed by setting $\Theta|_{z=z_h}=0$, where $\Theta$ is the expansion of outgoing null geodesics. A simple way to enforce this at all times during the numerical evolution is to impose a diffusion-like equation of the form~\cite{Attems:2017zam,Bea:2022mfb}
\begin{equation}
\label{eq:evolvgauge}
\left(\partial_v  \Theta+\kappa \,\Theta\right)|_{z=z_h}=0\,,
\end{equation}
with $\kappa>0$, so that the apparent horizon surface is pushed towards $z= z_h$ during the time evolution.\footnote{When $\kappa=0$ one recovers the strategy used in~\cite{Chesler:2013lia}. A disadvantage of that approach is that numerical errors inevitably accumulate, causing the apparent horizon surface to slowly drift away from the target surface $z=z_h$ during the evolution. This necessitates cumbersome regridding operations, which are avoided by using equation~\eqref{eq:evolvgauge} with $\kappa>0$.} This gives a second-order PDE for $\partial_v \xi$ of the form
\begin{align}
\label{eq:gauge}
    \left (A^\xi_{xx}\partial^2_x+A^\xi_{xy}\partial_x\partial_y+A^\xi_{yy}\partial^2_y+B^\xi_x\partial_x+B^\xi_y\partial_y+C^\xi \right)\partial_v\xi(v,x,y)=-S^\xi\,, 
\end{align}
where all the coefficients $A^\xi_{xx},A^\xi_{xy},A^\xi_{yy},B^\xi_x,B^\xi_y, C^\xi,S^\xi$ are known in terms of the functions $A,F_i,B,\Sigma,\theta$ and their derivatives; we have suppressed the explicit expressions for brevity. 
For more details, see appendix \ref{app:AH}.

\subsection{Boundary expansion}

In this work we are interested in analysing a specific deformation of the dual boundary field theory consisting of a non-trivial source, $\sigma_0$, for the energy-momentum tensor. Such a deformation is equivalent to considering the field theory not on three-dimensional Minkowski spacetime, but on the spacetime with line element
\begin{align}
ds^2=-dv^2+\sigma_0(v,x,y)^{2}\delta_{ij} dx^i dx^j\,.
\end{align}
As such, solving the Einstein equations order by order in $z$, we find that the asymptotic behaviour of our fields close to the AdS boundary is given by the expansion 
\begin{equation}
\label{eq:UVexp}
\begin{aligned}
A&=\frac{1}{z^2}+\frac{2}{z}\frac{(\xi-\partial_v\sigma_0)}{\sigma_0}+\left(-\frac{2}{\sigma_0}\partial_v \xi+\frac{\xi^2}{\sigma_0^2}+\frac{\partial_{x}^2\sigma_0+\partial_{y}^2\sigma_0}{\sigma_0^4}-\frac{\partial_{xx}\sigma_0+\partial_{yy}\sigma_0}{\sigma_0^3}\right)+ z a_3+\cdots,\\
F_i&= \partial_i\xi  - \frac{1}{\sigma_0^2}\left(-\partial_i\sigma_0\,\partial_v\sigma_0 + \sigma_0\,\partial_v\partial_i\sigma_0\right)+ z\, f_i\\
B&=z^3 b_3+\cdots\,,\\
\Sigma&=\frac{\sigma_0}{z}+\xi+\cdots\,,\\
\theta&=z^3 \theta_3+\cdots\,,\\
d_+B&=-\frac{3}{2}z^2 b_3+\cdots\,,\\
d_+\Sigma&=\frac{\sigma_0}{2z^2}+\frac{\xi}{z}+\left(\frac{\xi^2}{2\sigma_0^2}+\frac{\partial_{x}^2\sigma_0+\partial_{y}^2\sigma_0}{2\sigma_0^3}-\frac{\partial_{xx}\sigma_0+\partial_{yy}\sigma_0}{2\sigma_0^2}\right)+\frac{\sigma_0}{2}z a_3+\cdots\,,\\
d_+\theta&=-\frac{3}{2}z^2 \theta_3+\cdots\,,
\end{aligned}
\end{equation}
together with the constraints 
\begin{equation}
\label{eq:evol_boundary}
\begin{aligned}
\partial_v a_3&=\frac{1}{2 \sigma_0^8} \Big[ 20 (\partial_y \sigma_0)^4+20 (\partial_x \sigma_0)^4-16 (\partial_x \sigma_0)^2 (\partial_{y}^2 \sigma_0+2 \partial_{x}^2 \sigma_0) \sigma_0 \\
&+(5 (\partial_{y}^2 \sigma_0)^2+4 (\partial_{xy} \sigma_0)^2+6 \partial_{y}^2 \sigma_0 \partial_{x}^2 \sigma_0+5 (\partial_{x}^2 \sigma_0)^2+8 \partial_x \sigma_0 (\partial_{x}\partial_{y}^2 \sigma_0+\partial_{x}^3 \sigma_0)) \sigma_0^2\\
&-(\partial_{y}^4 \sigma_0+2 \partial_{x}^2\partial_{y}^2 \sigma_0+\partial_{x}^4 \sigma_0) \sigma_0^3-3 (\partial_y f_2+\partial_x f_1) \sigma_0^6 \\
&-6 a_3 \partial_v \sigma_0 \sigma_0^7+-8 (\partial_y \sigma_0)^2 (-5 (\partial_x \sigma_0)^2+2 (2 \partial_{y}^2 \sigma_0+\partial_{x}^2 \sigma_0) \sigma_0)\\
&+8 \partial_y \sigma_0 \sigma_0 (-4 \partial_x \sigma_0 \partial_{xy} \sigma_0+(\partial_{y}^3 \sigma_0+\partial_{x}^2\partial_y \sigma_0) \sigma_0) \Big]\,, \\
\partial_v f_1&=\frac{1}{3 \sigma_0}(6 \theta_3 \partial_y \sigma_0-6 (f_1 \partial_v \sigma_0+b_3 \partial_x \sigma_0)+(3 \partial_y \theta_3-\partial_x a_3+3 \partial_x b_3) \sigma_0),\,\\
\partial_v f_2&=\frac{1}{3 \sigma_0}(-6 f_2 \partial_v\sigma_0+6 b_3 \partial_y \sigma_0+6 \theta_3 \partial_x \sigma_0+(3\partial_x \theta_3 -\partial_y a_3-3\partial_y b_3)\sigma_0).
\end{aligned}
\end{equation}

In this expansion the coefficients $a_3,b_3, \theta_3, f_1, f_2$ correspond to expectation values, $\xi$ corresponds to the gauge fixing introduced in Section~\ref{subsec:gauge} and $\sigma_0(v,x,y)$ is a source for the breathing mode $\Sigma$ as seen by the leading behaviour of the metric close to the boundary.\footnote{Note that, under a Weyl transformation  $z=\tilde z \sigma_0$, the bulk metric asymptotes to 
\begin{align*}
ds^2=\frac{1}{\tilde z^2}\left(-\sigma_0(v,x,y)^{-2}dv^2+d\tilde z^2+\delta_{ij} dx^i dx^j\right)\,
\end{align*}
as $\tilde z\to0$, where $\sigma_0$ can now be interpreted as a tidal deformation.}

Furthermore, note that the constraints above are equivalent to demanding that the boundary stress tensor, defined through
\begin{align}
&T_{\mu\nu}=\lim_{z\to 0}\frac{2}{z}\left(K_{\mu\nu}-(K+2)\gamma_{\mu\nu}-(R_{\mu\nu}-\frac{1}{2}R\gamma_{\mu\nu})\right)\,
\end{align}
is covariantly constant;
here  $\gamma_{\mu\nu}$ is the induced metric on the boundary, $K_{\mu\nu}$ is the associated extrinsic curvature
and $R_{\mu\nu}$ is the three-dimensional Ricci tensor~\cite{Balasubramanian_1999}.
Using the asymptotic expansion \eqref{eq:UVexp}, one obtains the boundary stress tensor
\begin{equation}
\label{eq:boundary_stress_tensor}
\begin{aligned}
    T_{vv}&=a_3\\
    T_{vx}&=-\frac{1}{2 \sigma_0 ^5}(3 f_1 \sigma_0 ^5+4 \partial _{x}\sigma_0{}^3-\sigma_0  (\partial
   _{x}\sigma_0 (5 \partial _{xx}\sigma_0+3 \partial _{yy}\sigma_0
   )+2 \partial _{xy}\sigma_0 \partial _{y}\sigma_0)+4 \partial
   _{x}\sigma_0 \partial _{y}\sigma_0{}^2\\
   &+\sigma_0 ^2 (\partial
   _{xxx}\sigma_0+\partial _{xyy}\sigma_0))\\
   T_{vy}&=-\frac{1}{2 \sigma_0 ^5}(3 f_2 \sigma_0 ^5+\partial _{y}\sigma_0 (4 \partial _{x}\sigma_0
   {}^2-\sigma_0  (3 \partial _{xx}\sigma_0+5 \partial _{yy}\sigma_0))+\sigma_0  (\sigma_0  (\partial _{xxy}\sigma_0+\partial
   _{yyy}\sigma_0)\\
   &-2 \partial _{x}\sigma_0 \partial _{xy}\sigma_0  )+4 \partial _{y}\sigma_0{}^3)\\
   T_{xx}&=\frac{1}{2 \sigma_0 ^3}(a_3 \sigma_0 ^5+3 b_3 \sigma_0 ^5-4 \partial _{v}\sigma_0 \partial _{x}\sigma_0
   {}^2+\sigma_0  \partial _{v}\sigma_0 \partial _{xx}\sigma_0+4 \partial
   _{v}\sigma_0 \partial _{y}\sigma_0{}^2-\sigma  \partial _{v}\sigma_0
   \partial _{yy}\sigma_0\\
   &+4 \sigma_0  \partial _{vx}\sigma \partial
   _{x}\sigma_0-\sigma_0 ^2 \partial _{vxx}\sigma_0-4 \sigma_0  \partial
   _{vy}\sigma_0 \partial _{y}\sigma_0+\sigma_0 ^2 \partial _{vyy}(\sigma_0
   ))\\
   T_{yy}&=\frac{1}{2 \sigma_0 ^3}(a_3 \sigma_0 ^5-3 b_3 \sigma_0 ^5+4 \partial _{v}\sigma_0 \partial _{x}\sigma_0
   {}^2-\sigma_0  \partial _{v}\sigma_0 \partial _{xx}\sigma_0-4 \partial
   _{v}\sigma_0 \partial _{y}\sigma_0{}^2+\sigma_0  \partial _{v}\sigma_0
   \partial _{yy}\sigma_0\\
   &-4 \sigma_0  \partial _{vx}\sigma_0 \partial
   _{x}\sigma_0+\sigma_0 ^2 \partial _{vxx}\sigma_0+4 \sigma_0  \partial
   _{vy}\sigma_0 \partial _{y}\sigma_0-\sigma_0 ^2 \partial _{vyy}\sigma_0
   )\\
   T_{xy}&=\frac{1}{2 \sigma_0 ^3}(-3 \theta_3 \sigma_0 ^5+\partial _{y}\sigma_0 (4 \sigma_0  \partial
   _{vx}\sigma_0-8 \partial _{v}\sigma_0 \partial _{x}\sigma_0) \\
   &{}+2
   \sigma_0  (\partial _{v}\sigma_0 \partial _{xy}\sigma_0-\sigma_0  \partial
   _{vxy}\sigma_0+2 \partial _{vy}\sigma_0 \partial _{x}\sigma_0))
\end{aligned}
\end{equation}

\subsection{Field redefinitions}

For the numerical implementation, it is convenient to decompose the radial grid into two regions: an inner grid near the AdS boundary where boundary conditions are imposed and field theory observables are extracted, and an outer grid extending into the deep bulk. As seen in the asymptotic expansion above, several metric functions either diverge or vanish at the AdS boundary. We therefore introduce field redefinitions in the inner grid motivated by the asymptotic behaviour of these functions, so that the resulting variables remain finite and of order unity near the boundary. In the outer grid, we instead adopt simpler redefinitions, which prove advantageous for the equation used to fix the gauge variable $\xi$. Denoting by the subscripts $g1, g2$ the variables defined on the inner and outer grids respectively, the redefinitions we employ are as follows 
\begin{equation}
\begin{aligned}
\label{eq:fieldRedefinition}
A&=\frac{1}{z^2}+\frac{2}{z}\frac{(\xi-\partial_v\sigma_0)}{\sigma_0} - \frac{2}{\sigma_0}\partial_v \xi+\frac{\xi^2}{\sigma_0^2}+\frac{\partial_{x}^2\sigma_0+\partial_{y}^2\sigma_0}{\sigma_0^4}-\frac{\partial_{xx}\sigma_0+\partial_{yy}\sigma_0}{\sigma_0^3} + z A_{g1} \\
&=-\frac{2}{\sigma_0}\partial_v \xi+A_{g2}\,,\\
F_i&=\partial_i \xi+z F_{ig1}-\frac{1}{\sigma_0^2}\left(-\partial_i\sigma_0\partial_v\sigma_0+\sigma_0\partial_v\partial_i\sigma_0\right)\,=\partial_i \xi+F_{ig2}\,,\\
B&=z^3 B_{g1}\,=B_{g2}\,,\\
\Sigma&=\frac{\sigma_0}{z}+\sigma_0\xi+z^2\Sigma_{g1}+\partial_v\sigma_0\,=\Sigma_{g2}\,,\\
\theta&=z^3 \theta_{g1}\,=\theta_{g2}\,\\
%
%
d_+B&=z^2 d_+B_{g1}=d_+B_{g2}\,,\\
d_+\Sigma&=\frac{\sigma_0}{2z^2}+\frac{\xi}{z}+\left(\frac{\xi^2}{2\sigma_0^2}+\frac{\partial_{x}^2\sigma_0+\partial_{y}^2\sigma_0}{2\sigma_0^3}-\frac{\partial_{xx}\sigma_0+\partial_{yy}\sigma_0}{2\sigma_0^2}\right)+z d_+\Sigma_{g1}=d_+\Sigma_{g2} \,,\\
d_+\theta&=z^2 d_+\theta_{g1}=d_+\theta_{g2}\,.
\end{aligned}
\end{equation}
The boundary conditions for the new fields at the edge of the inner grid are then simply
\begin{align*}
A_{g1}&=a_3\,,\\
F_{ig1}&=f_i\,,\\
B_{g1}&=b_{3}\,,\\
\Sigma_{g1}&=0\,,\\
\theta_{g1}&=\theta_{3}\,\\
d_+B_{g1}&=-\frac{3}{2}b_{3}\,,\\
d_+\Sigma_{g1}&=\frac{\sigma_0}{2} a_3\,,\\
d_+\theta_{g1}&=-\frac{3}{2}\theta_{3}\,.
\end{align*}
At the interface between the two grids, we demand continuity of the fields  and their first radial derivative.

\section{Numerical computation}
\label{sec:numerics}

We now give some details about the numerical evolution procedure. As mentioned, this is done using the infrastructure Jecco, originally introduced in~\cite{Bea:2022mfb}, and which was modified as follows.

Originally, Jecco was developed to perform 3+1 time evolutions in a five-dimensional spacetime, with a constant source for the scalar on the AdS boundary. In this work we aim to study various aspects of gravitational turbulence in a purely gravitational context, in a setup where the boundary is 2+1 dimensional. As such, we have modified the original code to carry out 3+1 time evolutions in an asymptotically AdS$_4$ spacetime; we have removed the scalar field, and instead implemented a non-trivial source in the metric with a dependence on $v,x,y$.

\subsection{Evolution algorithm}
\label{subsec:evolution_algorithm}

The system of equations~\eqref{eq:eqns} and~\eqref{eq:gauge} has a natural nested structure which we will exploit in order to construct the solutions we are after. In particular, we will follow the protocol laid down below:\\

\noindent On the initial time slice, $v=0$ ($n=0$), the procedure consists of
\begin{enumerate}
    \item Giving $B(0,z,x,y),$ $ \theta(0,z,x,y), a_3(0, x,y), f_i(0, x,y)$ according to our initial data (see Sec.~\ref{sec:S_ID_TW}) and an initial guess for $\xi_\mathrm{init}=\xi(0, x,y)$.
    \item Solving successively the ODEs~\eqref{eq:eqns} for $\Sigma, F_i, d_+\Sigma,(d_+B, d_+\theta), A$; note that, as a result of the Bianchi identities, the equations for $d_+F_i,d_+d_+\Sigma$ are automatically satisfied. Essentially here we solve the corresponding ODEs in $z$ for every point in the $(x,y)$ plane, subject to boundary conditions.
    \item Solving iteratively equation \eqref{eq:C7} in order to locate the apparent horizon surface, $H(x,y)$, and adjust $\xi$ accordingly, $\xi_\mathrm{init } \leftarrow \xi_\mathrm{init}+(z_h-H(x,y))$, in order to place it at a constant radius $z=z_h$. See appendix~\ref{app:AH} for more details.
    \item Given the updated profile for $\xi$ and the initial data for $B(0,z,x,y), \theta(0,z,x,y)$, $ a_3(0, x,y), f_i(0, x,y)$, we solve successively the ODEs for the remaining fields, as in step 2.
    \item Having these at hand, we compute $\partial_v B,\,\partial_v \theta$ through the definition~\eqref{eq:d+f}, $\partial_v \xi$ through~\eqref{eq:gauge}, and $\partial_v a_3,\, \partial_v f_i$  through \eqref{eq:evol_boundary}, and carry out a time step in order to find $B(v_{1},z,x,y),\theta(v_{1},z,x,y),$ $ \xi(v_{1}, z,x,y),$ $ a_3(v_{1},  x,y)$ and $f_i(v_{1},  x,y)$. 
\end{enumerate}
On slices with $n\geq 1$, we proceed as follows
\begin{enumerate}
\item at any given time $v_n$ we know $B(v_n,z,x,y),$ $ \theta(v_n,z,x,y),$ $\xi(v_n, x,y), a_3(v_n, x,y)$ and $f_i(v_n, x,y)$;
\item successively solve the elliptic equations for $\Sigma, F_i, d_+\Sigma,(d_+B, d_+\theta), A$, subject to boundary conditions;
\item carry out a time step in order to find $B(v_{n+1},z,x,y),\theta(v_{n+1},z,x,y),$ $ \xi(v_{n+1}, z,x,y),$ $ a_3(v_{n+1},  x,y)$ and $f_i(v_{n+1},  x,y)$ as in point 5 of the scheme above.
\item repeat.
\end{enumerate}

As mentioned earlier, the numerical integration is performed in two coordinate patches: the patch $g1$ close to the boundary (inner) and the patch $g2$ close to the horizon (outer). For convenience in each patch we are solving for slightly different fields, which are explicitly defined in \eqref{eq:fieldRedefinition}. At the boundary interface between the inner and outer grid we impose continuity of the fields and their first derivative.

We discretise the directions $x,y$ using a uniform grid in the domain $[0,L_x]\times [0,L_y]$ and we approximate the corresponding derivative operators using fourth order finite difference, subject to periodic boundary conditions. To damp high-frequency noise we add a sixth-order Kreiss-Oliger numerical dissipation term. This is done by replacing
\begin{equation}
f \leftarrow f + \eta \frac{h^6}{64} \left(\partial_{x}^6+\partial_{y}^6 \right) f \,,
\end{equation}
for all evolved quantities $f$. Here $h$ is the grid spacing and $\eta$ is a tunable dissipation parameter which for stability it should be less than one; unless otherwise stated, we use $\eta=0.2$.
Note that this term approaches zero faster than the error in the spatial finite difference as $h\to0$. For the radial direction $z$, 
pseudo-spectral collocation methods are used to approximate the derivatives in a Chebyshev-Lobatto grid; see~\cite{Bea:2022mfb} for implementation details. For the time stepping, we use an Adams-Bashforth integrator.

In appendix \ref{app:tests} we discuss several tests that were carried out to verify that the code works as expected.

\subsection{Initial data and source: turbulent wake}
\label{sec:S_ID_TW}

According to the evolution algorithm described above, in order to carry out the simulations of interest we need to specify the initial data $B(0,z,x,y),$ $ \theta(0,z,x,y),\xi(0, x,y),$ $a_3(0, x,y)$, $f_i(0, x,y)$. In the case of turbulent wakes, in our numerical experiments we consider initial data corresponding to a boosted black brane (which has the dual interpretation of a fluid moving with four velocity $U^\mu$ and temperature $T$) with a line element given by
\begin{equation}
\begin{aligned}
ds^2 & =\frac{1}{\rho^2}\left[2 U_m dx^m d\rho-f(b\rho)U_m U_n dx^m dx^n+ P_{m n}dx^m dx^n\right]\,,\\
P_{mn} &=\eta_{mn}+U_m U_n\,,\\
f(\rho) &=1-\rho^3\,.
\end{aligned}
\end{equation}
In these coordinates the boundary is at $\rho=0$ and the horizon at $\rho_h=1/b=3/(4\pi T)$, with the velocity field constrained by $U^mU_m=-1$. To bring this to the form of the metric
used for the evolution, we perform a coordinate transformation $x^m=y^m+\zeta^m(\rho)$, where the vector $\zeta$ is chosen such that the 
 $g_{i\rho}, g_{\rho,\rho}$ components of the transformed metric vanish. We obtain~\cite{Balasubramanian:2013yqa}
\begin{align}
ds^2=\frac{1}{\rho^2} \left[\left(\eta_{mn}+(b\rho)^3 U_m U_n \right)dy^{m} dy^{n}-\frac{2}{\sqrt{f(b\rho)+(U^0)^2(b\rho)^3}}dv\,d\rho \right]\,.
\end{align}
Comparing this with the metric ansatz introduced in~\eqref{eq:ansatz}, we get
\begin{equation}
\label{eq:match}
\begin{aligned}
F_i&=\frac{(b\rho)^3}{\rho^2} U_0U_i \,,\\
\sinh\theta&=\frac{(b\rho)^3 U_xU_y}{\sqrt{1+(b\rho)^3(U_x^2+U_y^2)}}\,,\\
A&=\frac{1}{\rho^2}\left[1-(b\rho)^3(1+U_x^2+U_y^2) \right]\,,\\
\Sigma^2 \rho^2&=\sqrt{1+(b\rho)^3(U_x^2+U_y^2)}\,,\\
e^{2B}&=\frac{1+(b\rho)^3 U_x^2}{1+(b\rho)^3U_y^2} \,,
\end{aligned}
\end{equation}
together with the differential equation
\begin{align}
\label{eq:rtorho}
\frac{dr}{d\rho}=-\frac{1}{\rho^2}\frac{1}{\sqrt{1+(b\rho)^3(U_x^2+U_y^2)}}\,,
\end{align}
which is in agreement with \cite{Adams:2013vsa}.  The values of $a_3(0,x,y), f_i (0,x,y)$ are extracted  from the asymptotic behaviour of these functions as $z=1/r\to 0$. Without loss of generality, we set initial data with $U_y=0$ and we treat $U_x$ and $b$ (or, equivalently, the temperature $T$) as free parameters. %

We now move on to discuss the source used in our turbulent wake simulations. Specifically, the source $\sigma_0(v,x,y)$ is treated as a Gaussian-type quench with 
\begin{equation}
\sigma_0(v,x,y)=1-\frac{A}{2}\left(1+\tanh{\frac{v-v_0}{\tau}}\right)\exp\left[ \frac{-(x-x_s)^2}{2 (\sigma_{x} L_x)^2} \right] \exp\left[ \frac{-(y-y_s)^2}{2 (\sigma_y L_y)^2} \right],
\label{eq:source_profile}
\end{equation}
where $(x_s,y_s)$  and $(\sigma_x,\sigma_y)$ parametrise the location and the size of the Gaussian, $A$ is a fixed overall amplitude controlling the strength of the forcing, $v_0$ controls the time when the source is turned on and $\tau$ how fast it is turned on. 
Note that at $v=0$, $\sigma_0(v,x,y)\sim1$.

\section{Results}
\label{sec:results}
 In this section we discuss our results for turbulent wakes. We set initial conditions corresponding to a homogeneously boosted black brane with velocity $U_x$ flowing from left to right ($U_y = 0$), and introduce a localised deformation by quenching $\sigma_0$ to a symmetric Gaussian profile centred at $(x_s, y_s) = (0, 0)$, as described in Section~\ref{sec:S_ID_TW} (see equation~\eqref{eq:source_profile}). Unless otherwise stated, all results presented in this section correspond to the parameter choices $L_x = 60000$, $L_y = 10000$, $U_x = 0.3$, $z_h = 6$, $\sigma_x = \sigma_y = \frac{L_x}{24\pi}$. Care needs to be taken when selecting the value of $v_0$ and $\tau$ in  equation~\eqref{eq:source_profile}, which control the timescale and the speed of the source activation. If the activation occurs too soon or too rapidly, an interference pattern propagates through the whole domain and contaminates the physical solution, making it prohibitive for the study of turbulent wakes. For this reason, we consider a slow, adiabatic activation with $v_0=0.25 \frac{L_x}{U_x}$ and $\tau=0.4 v_0=0.1 \frac{L_x}{U_x}$ 
 , for which such interference is absent. In terms of the numerics, we discretise the spatial directions using a uniform grid with $N_x = 450$ and $N_y = 75$ points, with one inner and one outer radial domains of $N_z = 10$ Chebyshev points respectively, and Kreiss--Oliger dissipation parameter $\eta = 0.2$. This constitutes our reference simulation; a brief discussion on the parameter space is presented at the end of this section. With this setup at hand, we proceed to analyse both boundary observables and geometric quantities defined purely in terms of bulk gravitational variables. Note that, since this configuration is neither statistically homogeneous nor isotropic, we do not expect the flow to exhibit power-law scaling in the energy spectrum; rather, the relevant signatures of turbulent wake dynamics are the self-similar profiles and scaling relations discussed below.

In analysing our simulation, we will draw intuition from turbulence in fluids, and as such, we find it useful to define the notion of boundary velocity. In the literature, there are two ways to do so. On the one hand, following \cite{Andrade:2019rpn}, we can define the velocity 
\begin{equation}
u^i = (f^i-\nabla^i a_3)/a_3\,;
\end{equation}
this expression is motivated by \cite{Andrade:2018zeb}, which showed that, given this, the Einstein equations in the large-$D$ limit reduce to the conservation equation of the exact second-order Landau-frame hydrodynamic stress tensor. On the other hand, we can define the velocity through the boundary stress-energy tensor. Specifically, \cite{Adams:2013vsa} defined the velocity $u_i$ as the normalised future-directed time-like eigenvector of $\langle T^i{}_{j}\rangle$, namely
\begin{equation}
\langle T^i{}_{j}\rangle u^j=-\epsilon u^i
\end{equation}
where $\epsilon$ is the eigenvalue representing the proper energy density.

\begin{figure}
    \centering
    \includegraphics[width=0.45\linewidth]{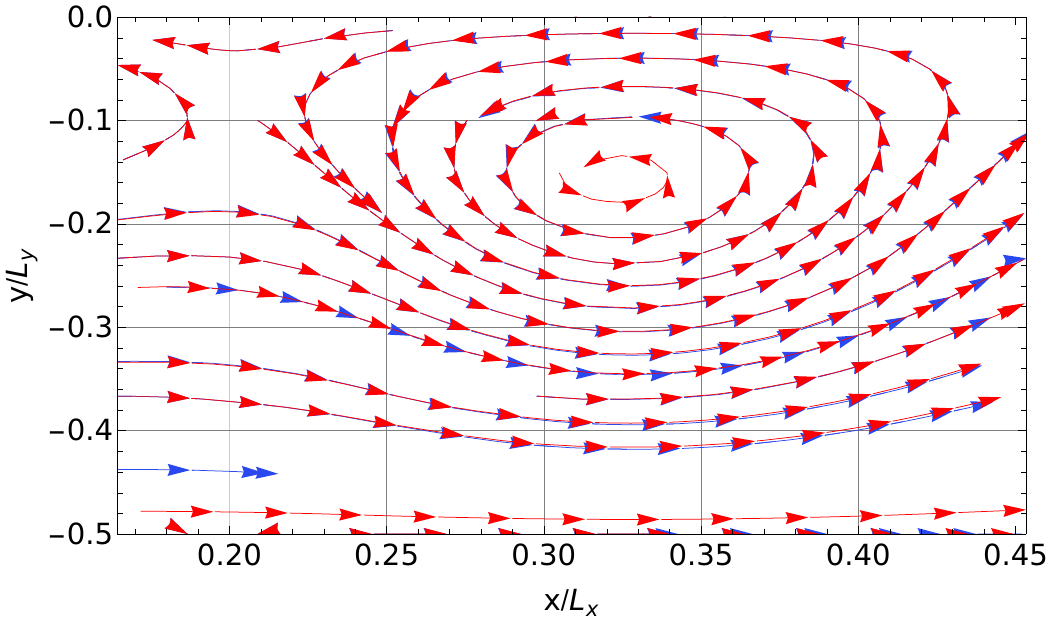}
    \includegraphics[width=0.45\linewidth]{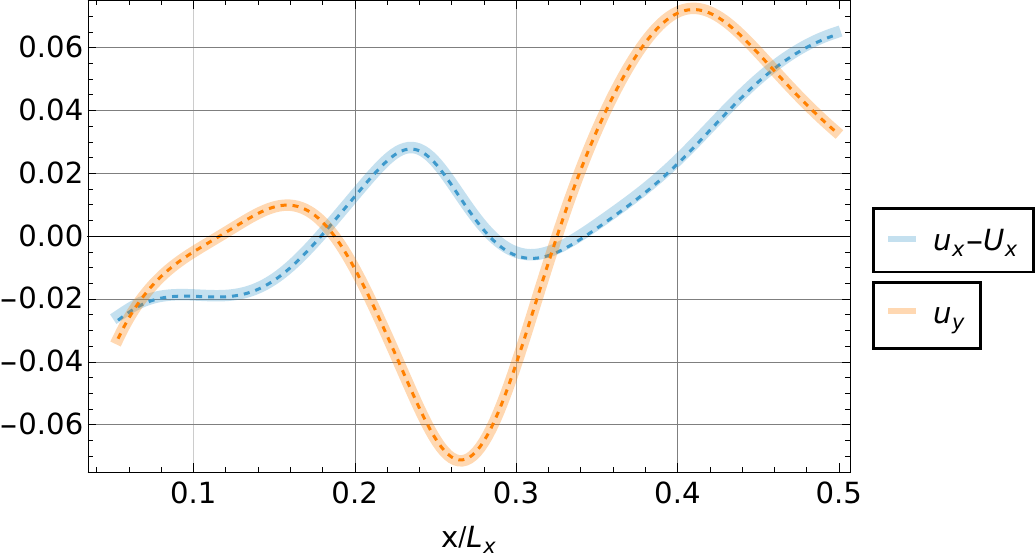}
    \caption{(Left) We present a qualitative comparison for a gravitational wake simulation by overlaying a snapshot of the streamlines of the velocity vector fields computed using two different formulations, focusing on a region containing a vortex. (Right) Plot of the velocity deficit in the $x$ direction, $u_x-U_x$, and  $u_y$ as functions of $x$ for $y/L_y=0.2466$, $v/v_0=8.05$ for the same region as in the left panel. The two plots together indicate that the formation and evolution of the flow structures can be reliably identified using either formulation of the velocity field.}
    \label{fig:velocities_diff}
\end{figure}

In the left panel of Fig.~\ref{fig:velocities_diff} we overlay a snapshot of the velocity field at $v=8.05 v_0$ computed via the two different definitions for a turbulent wake with parameters defined above, focusing on a region containing a vortex. We see that the two quantities behave qualitatively the same. In the right panel of Fig.~\ref{fig:velocities_diff} we show a quantitative comparison of the two. Specifically, we plot $u_x(v_a, x, y_a)-U_x$ and  $u_y(v_a, x, y_a)$, where $U_x$ is the background velocity associated with the black brane initial data, as function of $x$ for $v_a=8.05 v_0$ and $y_a=0.2466 L_y$ fixed. Modulo the region around the source, we see good quantitative agreement between the two approaches, indicating that the formation and evolution of the flow structures can be reliably identified using either velocity field. We have checked that, close to the source, the matrix $\langle T^i{}_{j}\rangle$ has a large condition number, making the extraction of its eigenvalues more prone to numerical error. Going forward, we will use the first definition as it is computationally easier. 

\begin{figure}[thpb]
     \centering
     \includegraphics[width=0.98\textwidth]{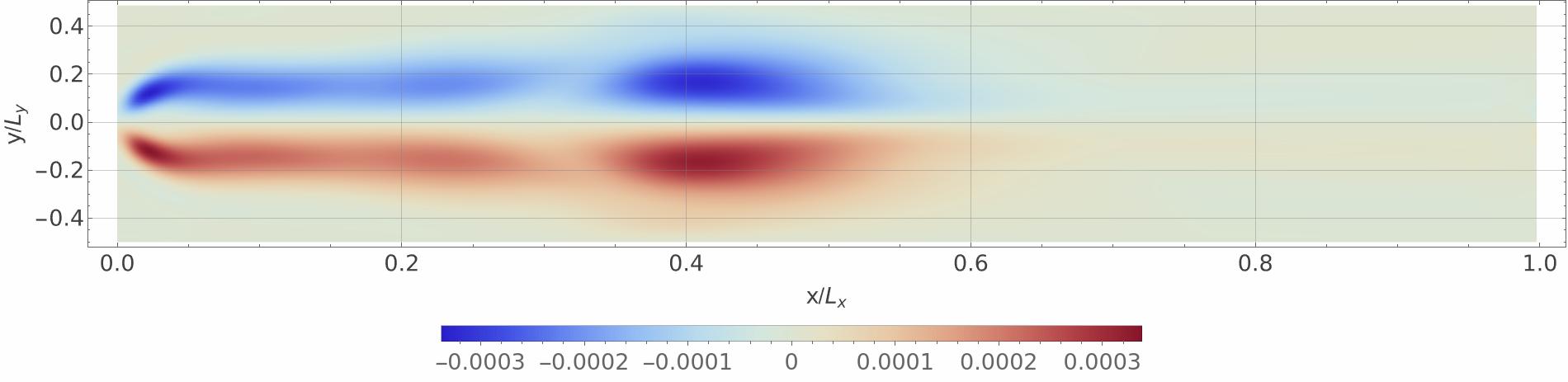}
    \\
     \includegraphics[width=0.98\textwidth]{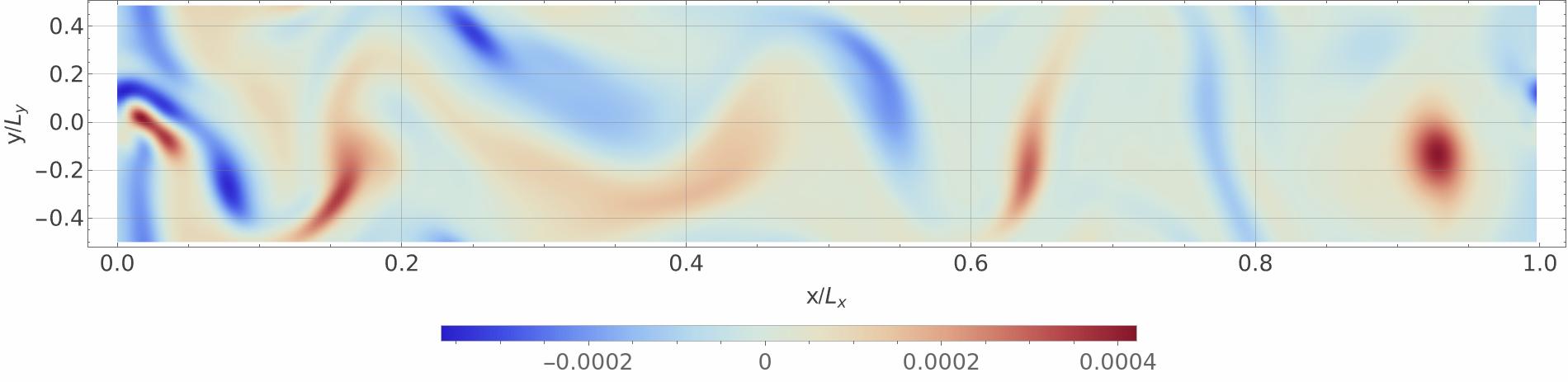}
    \caption{Snapshots of the boundary vorticity at different times. The source is located on the left, at $(x_s,y_s)=(0,0)$, and the background flow goes from left to right. Clusters can be interpreted as vortices and antivortices regions. One can observe how the symmetric propagation of the vortices breaks in a chaotic wake dynamic from the top panel to the bottom panel.
    \label{fig:Turbulent_wake_u0d3_LX60000}
    }
\end{figure}

\begin{figure}[htb]
    \centering
    \includegraphics[width=0.75\linewidth]{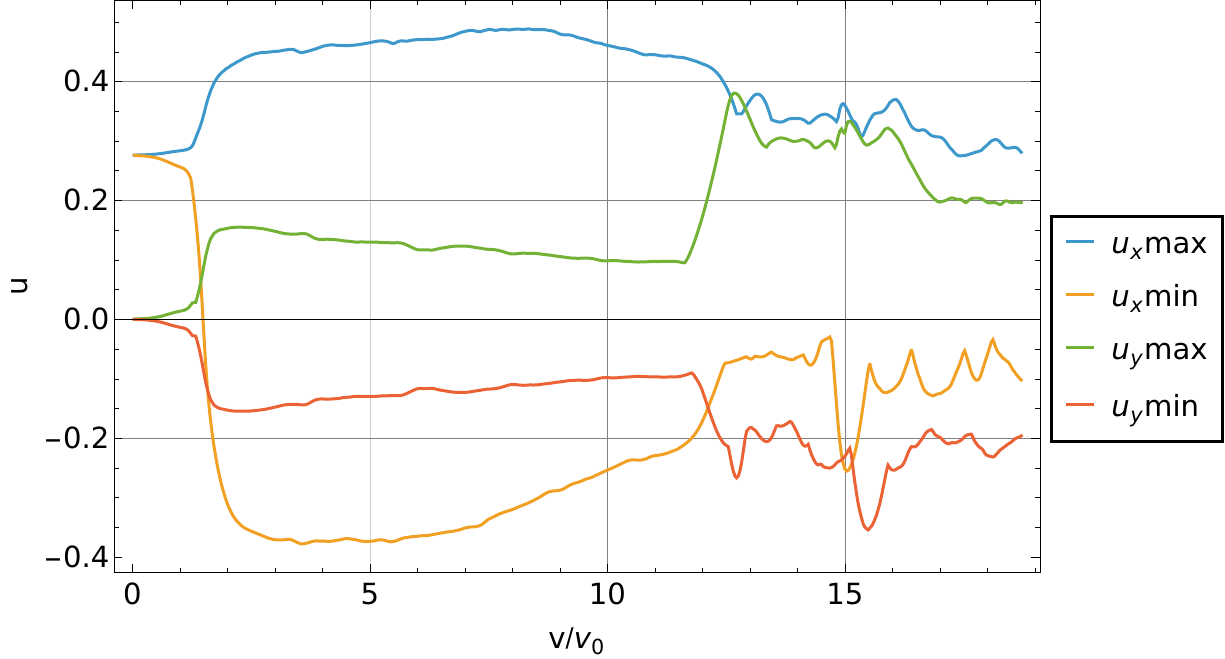}
    \caption{Evolution of the minimum and maximum values for $u_x$ and $u_y$. The first phase, up to $v/v_0\sim2$, corresponds to the source activation. The second phase, where $u_y$ exhibits $y\to-y$ symmetry, corresponds to the symmetric phase. The third phase, starting around $v/v_0=12$, exhibits breaking of the $y\to-y$ symmetry, and the system exhibits chaotic dynamics.}
    \label{fig:velocities_extrema_u0d3}
\end{figure}

Fig.~\ref{fig:Turbulent_wake_u0d3_LX60000} shows fixed-time snapshots of the vorticity for our reference simulation. As the source is switched on, we observe the formation of pairs of vortices behind the source, subsequently detaching and being advected downstream by the background flow. The evolution naturally splits into three phases, clearly visible in Fig.~\ref{fig:velocities_extrema_u0d3} showing the extrema of $u_x$ and $u_y$:

\begin{enumerate}
    \item A brief initial phase, where the source gets activated and the first pair of vortices appears. This phase is determined by the values of $v_0$ and $\tau$ defined in equation~\eqref{eq:source_profile}: larger values of $v_0$ and $\tau$ delay the activation of the source and thus, the first pair of vortices.
    \item The reflection-symmetric phase, in which the dynamics are approximately   invariant under $y\rightarrow -y$, resulting in the propagation of symmetric  vortex pairs downstream. We have observed that the time of detachment of the first pair of vortices is related to the source activation timescale~$\tau$.
   \item The final phase where the reflection symmetry is broken, leading to a more irregular, chaotic evolution. This phase is relatively short-lived in our simulations, as the wake returns to its origin and interacts with the source due to the periodic boundary conditions in $x$.
\end{enumerate}

We now proceed to analyse the turbulent wake using tools developed in fluid dynamics. Within this context, wake dynamics are studied with the hypothesis that the flow is statistically stationary, and hence all the quantities are time averages. Having said that, in our setup, averaging over time requires extra care. The main issue is the periodicity of the domain in the $x$ direction: given that the wake will return to the source once it crosses the $x$-boundary, our setup does not allow for a complete free shear flow condition (which requires an infinite domain).  This issue can be ameliorated by using a very large $x$ domain, but this is prohibitively expensive  computationally. 

In our studies, we implemented a time average that considers different time windows depending on the $x$ coordinate. The raw idea is to consider a time window starting from the first passage of the first pair of vortices from  a given location $x$, and ending in the second passage of those vortices from the same point, after they have crossed the periodic boundary; this allows us to consider long enough time intervals for each $x$, but still exclude times when the wake is spoiled by the periodic boundary conditions. 
To compute these time intervals, we determine (i) the time of arrival of the first pair of vortices to the specific $x$ location and (ii) the vortex propagation speed, computed directly from the simulation during the ``clean'' phase. The vortex propagation speed is extracted by looking for peaks in the transverse velocity. Indeed, the initial conditions are set such that $U_y=u_y=0$, so a peak would flag the arrival of a strong perturbation, in our case a vortex. Plotting the profile of this arrival time allows us to perform a simple linear fit and extract the velocity of the propagation. Using this velocity, we compute the time interval taken by the wake to travel a full domain length $L_x$. As discussed above, the starting time $v^{start}$ of each average will depend on $x$, but the time interval, $\Delta v^{aver}$, on which we average will be the same.

The time-averaged quantities discussed below are meaningful only if the averaging interval is long compared to the dynamical timescales of the vortex structures. To verify this, we compare the averaging interval against two complementary diagnostics:

First, we estimate the vortex turnover time by identifying a representative vortex and computing the mean angular velocity, averaged over all points within a distance $\rho_{\mathrm{fixed}}$ from the vortex centre. Since the angular velocity exhibits both radial and azimuthal variations, we repeat this procedure for different values of $\rho_{\mathrm{fixed}}\in[0.05, 0.25]\,L_y$, finding turnover times in the range  $[\Delta v^{\mathrm{turnover}}_{\min}=0.749783, \Delta v^{\mathrm{turnover}}_{\max}=1.0507]\,v_0$; the duration of the full averaging interval is $\Delta v^{aver}=13.3152\, v_0$. This gives a conservative lower bound of at least 12 averaging-interval to turnover-time ratios, indicating that the averaging window spans many dynamical timescales of the individual vortex structures.

Second, we count the number of vortex passages through a representative $x$-location during the averaging interval. We monitor the transverse velocity $u_y$ at a fixed $(x,y)$ position sufficiently far from the source: the passage of a vortex is identified by the characteristic signature of a positive peak, a zero crossing, and a negative peak (or the reverse, depending on the sense of rotation). Applying this procedure, we find that the averaging interval includes the passage of approximately 5 vortex pairs, each consisting of a vortex and its counter-rotating companion.

These two diagnostics are complementary: the turnover time measures how well each individual vortex is internally resolved by the averaging window, while the passage count measures how many statistically independent vortex events contribute to the average at each $x$-location. Taken together, they suggest that the averaging interval is sufficient for extracting first-order statistics such as the mean self-similar velocity profiles discussed below, even though fully converged higher-order statistics cannot be claimed. This conclusion is further corroborated by the large-$D$ analysis in Appendix~\ref{app:wake_largeD}, where a larger computational domain yields noticeably better satisfaction of the self-similarity constraint, consistent with domain size being the limiting factor.

As a direct check on the convergence of the time averages, in Table~\ref{tb:conv} we compare the changes in the norm of the mean velocity as the averaging window is progressively extended for a fixed value of $x$: specifically, for $x_a=0.38 L_x$, we list 
$C=|u_{x}^{\Delta v^{aver}_j}(x_a,y) - u_{x}^{\Delta v^{aver}_i}(x_a,y)|$; in this notation $u_{x}^{\Delta v^{aver}_i}(x,y)$ indicates the longitudinal velocity $u_x$ averaged on the time window $\Delta v^{aver}_i=v_i-v^{start}$, with $v_i < v_j$ and $v_{i+1}-v_i=const$. Note that, since we fix the $x$ location, $v^{start}$ is always the same. The successive differences decrease in magnitude, demonstrating that the  time-averaged profiles are converging as the averaging window is extended, and  that the mean self-similar profiles extracted over the full window are meaningful.

\begin{table}[h]
\centering
\begin{tabular}{|c|c|c|}
\hline
$v_i/v_0$ & $v_j/v_0$ & $C$ \\
\hline
9.1104  & 10.1616  & 0.0137673 \\
10.1616  & 11.2128  & 0.0109882 \\
11.2128  & 12.264  & 0.00816854 \\
12.264  & 13.3152  & 0.00756438 \\
\hline
\end{tabular}
\caption{ Convergence of the time-averaged longitudinal velocity $u_x$ as the averaging window is extended for $x_a=0.38 L_x$. We report the 
difference $C=|u_{x}^{\Delta v^{aver}_j}(x_a,y) - u_{x}^{\Delta v^{aver}_i}(x_a,y)|$ for successive 
equispaced extensions of the time interval. The monotonically decreasing values demonstrate 
that the time-averaged profiles converge as the averaging window is extended.}
\label{tb:conv}
\end{table}

\begin{figure}[thbp]
    \centering
    \includegraphics[width=1\textwidth]{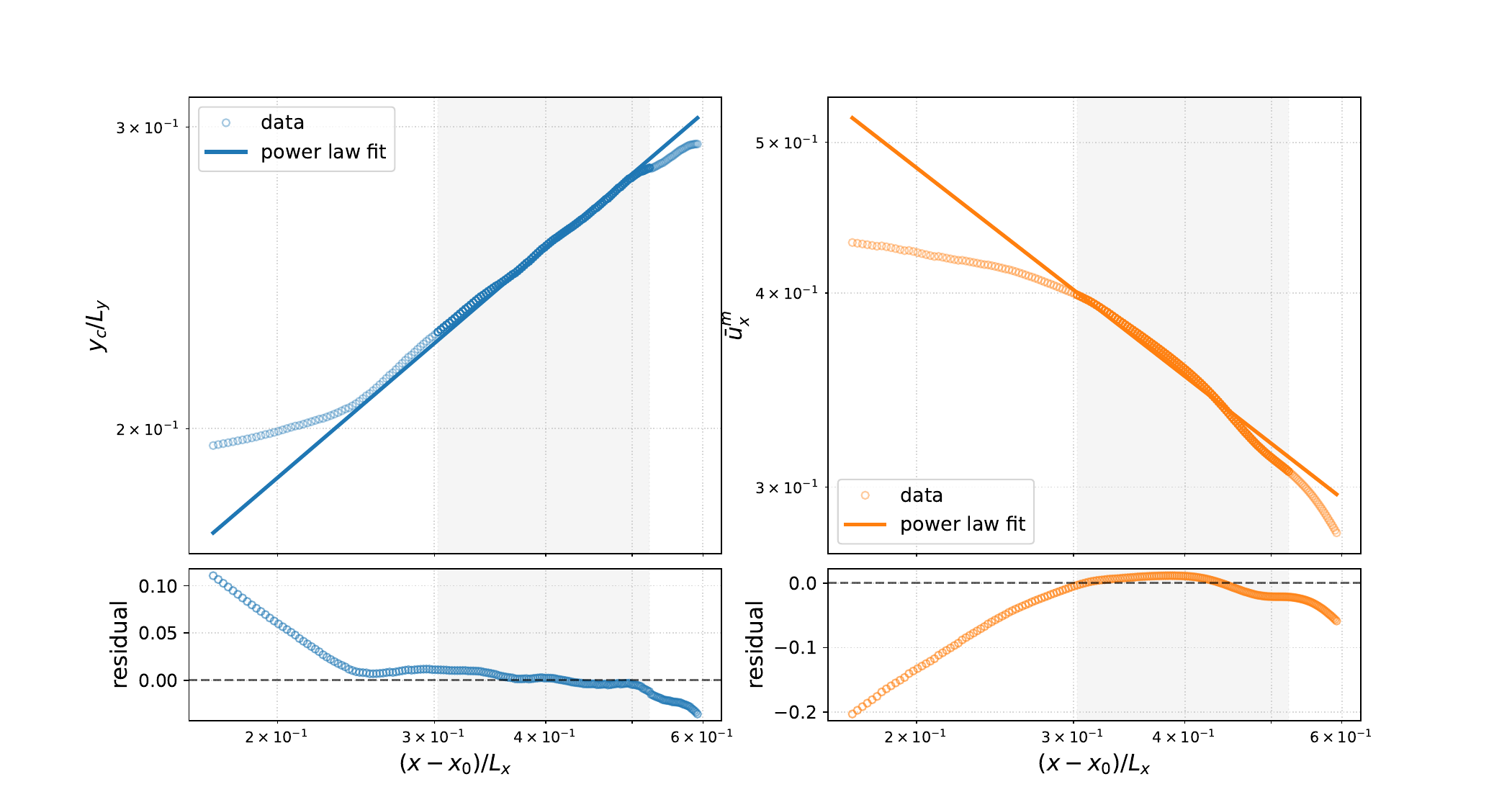}\\
    \includegraphics[width=0.5\textwidth]{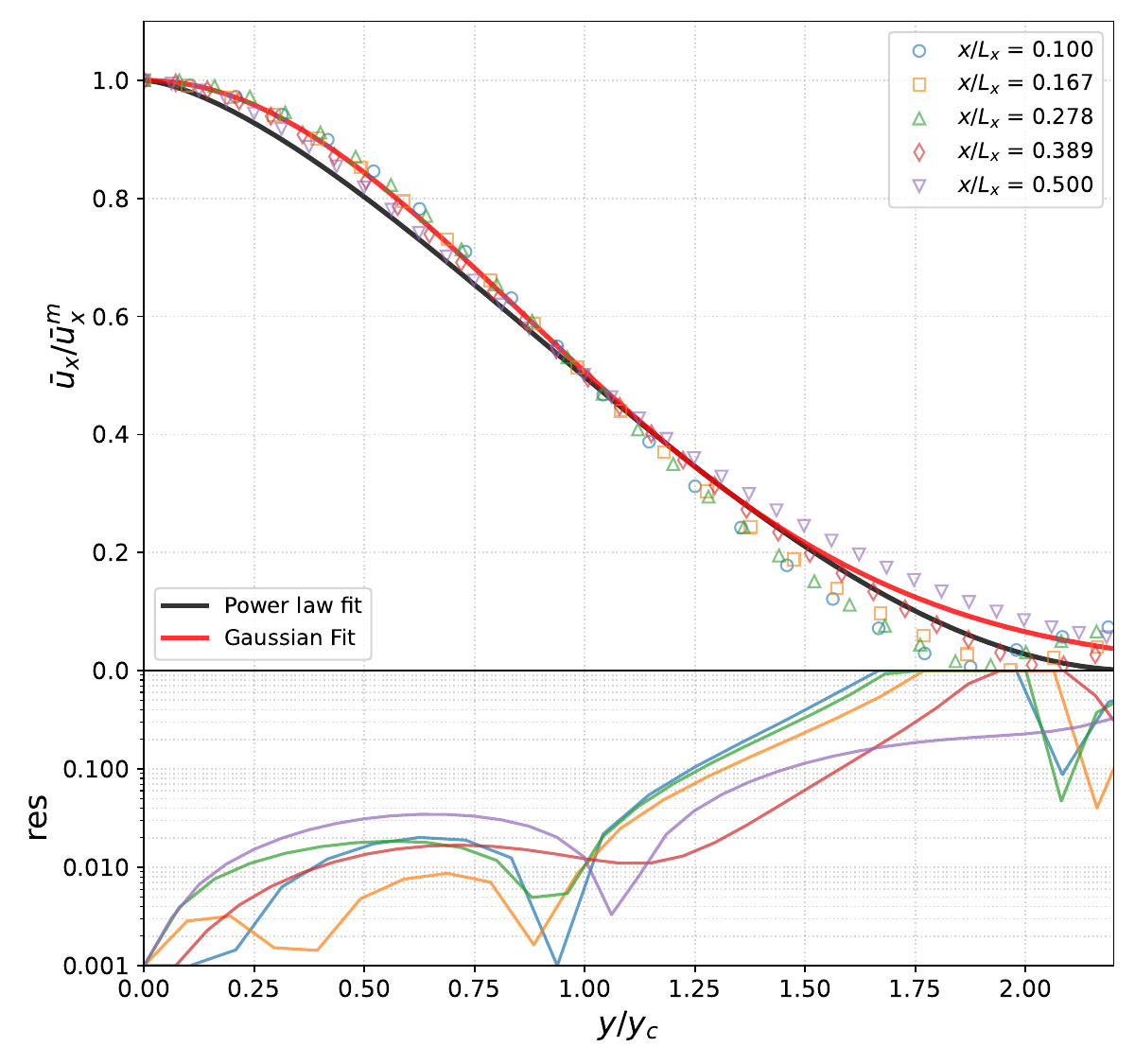}
         \caption{(Top, Left) The wake width, $y_c$, as a function of $x$, compared with the theoretical prediction. (Top, Right) The maximum of the velocity deficit, $\bar{u}_x^m$ as a function of the $x$, compared with the theoretical prediction. (Bottom) Self-similar profiles of the wake at distinct $x$ values, where $\bar{u}_x/\bar{u}_x^m$ is plotted against $y/y_c$. We plot only the positive $y/y_c$-semi axis since the profiles are mostly symmetric.}
         \label{fig:jecco_deficit}
\end{figure}

Following the fluid dynamics literature, we focus on analysing the maximum velocity deficit $\bar{u}_x^m$ defined as 
\begin{align}
    \bar u_x(x,y) &= U_x - u_x(x,y)\\
    \bar{u}_x^m(x) &= \max_y \bar u_x(x,y)
\end{align}
and the width of the wake $y_c$ defined implicitly via
\begin{align}
u_x(x,y_c) &= \frac{1}{2}\bar{u}_x^m(x)
\end{align}
where $U_x$ is the velocity of the boosted black brane in the background; note that the quantities above are time-averaged and hence have no $v$ dependence. In Fig.~\ref{fig:jecco_deficit} we plot these two quantities at different values of $x$. Motivated by the classical theory of self-similar planar turbulent wakes, we fit our data to
\begin{align}
    y_c(x) = A\,(x-x_0)^{a}, \qquad \bar{u}_x^m(x) = B\,(x-x_0)^{-a},
\end{align}
where $x_0$ accounts for the finite size of the source and is treated as a free parameter. For an incompressible flow, conservation of the wake momentum deficit,
\begin{equation}
  \int_{-L_y/2}^{L_y/2} \bar{u}_x(x,y)\, dy = \mathrm{const},
\end{equation}
implies $\bar{u}_x^m\cdot y_c= \mathrm{const}$, and making the additional assumption that turbulent mixing acts as an effective diffusion mechanism (see \cite{10.7551/mitpress/6781.001.0001} for details) yields $y_c\propto x^{1/2}$ and $\bar{u}_x^m\propto x^{-1/2}$, with the two exponents equal and opposite. In our case, the flow is compressible: the Mach number $\mathrm{Ma} = U_x/c_s \approx 0.42$ places us in a regime where compressible effects are present but not dominant\footnote{We note that~\cite{Waeber:2021xba} considers homogeneous and isotropic turbulence with forcing through the space-space part of the metric, finding a flow dominated by the incompressible component; the maximum of the velocity field was $0.035$, giving rise to a Mach number of $\sim 0.05$. Here, the Mach number is a full order of magnitude larger, which is compatible with us observing some compressibility effects.
}. In a compressible flow, density variations modify the momentum deficit integral and the constraint $\bar{u}_x^m \cdot y_c = \mathrm{const}$ no longer follows directly, meaning the two exponents need not be exactly equal and opposite. Consistently, we find that the momentum deficit integral decreases by approximately $8\%$ over the full evolution, confirming that incompressibility is only approximately satisfied. In a compressible flow, density variations modify the momentum deficit integral and the constraint $\bar{u}_x^m \cdot y_c = \mathrm{const}$ no longer follows directly, meaning the two exponents need not be exactly equal and opposite. However, since the momentum deficit integral decreases by only approximately $8\%$ over the full evolution, compressible corrections are small and the leading-order incompressible prediction is expected to hold to a good approximation. We therefore fit both profiles with the same exponent $a$, consistent with the incompressible theory at leading order, and expect $a$ to be close but not exactly equal to $1/2$. 

We find $a = 0.446\pm0.003$ and $x_0 \sim 1.13\pm0.006\,\sigma$ (where the error values refer to the $1\sigma$ marginal standard error), consistent with the expectation that the virtual origin of the wake is displaced from the source centre by a distance of order the source scale. The fitted exponent is in good agreement with the classical prediction $a = 1/2$, and the fits are good over the portion of the longitudinal domain where the free shear flow condition holds, as can be seen from the small residuals in Fig.~\ref{fig:jecco_deficit}. The deviation from the classical exponent is modest and likely reflects the finite domain size and the influence of the periodic boundary conditions, rather than a departure from self-similar wake phenomenology -- this is further supported by the self-similar analysis of turbulent wakes in the large-$D$ limit presented in Appendix~\ref{app:wake_largeD}. The fits are valid only within a certain interval in $x$: for the width $y_c$, the expansion of the wake eventually reaches the periodic boundary of the $y$-domain and spoils the natural transverse growth, while the fit for $\bar{u}_x^m$ holds until the $x$-boundary is probed and the free shear flow condition is broken.

In the bottom panel of Fig.~\ref{fig:jecco_deficit} we present $\bar{u}_x(x,y)/\bar{u}_x^m(x)$ as a function of $y/y_c$ for different values of $x$, observing a self-similar structure.  In more detail, in the self-similar regime the normalised velocity deficit is expected to collapse onto a universal profile,
\begin{equation*}
    \frac{\bar{u}_x}{\bar{u}_x^m}=f\!\left(\frac{y}{y_c}\right).
\end{equation*}
This is precisely what we observe in the figure; note that this  also provides an independent spatial consistency check on the quality of the time averages. To characterise this profile, we compare our data with two classical forms from turbulent wake theory; for details on the origin of these forms see \cite{10.7551/mitpress/6781.001.0001,Pope_2000}. 
Specifically, we use the Gaussian profile 
\begin{equation}\label{eq:gaussSS}
    \frac{\bar u_x}{\bar{u}_x^m} = \exp\!\left[-C_2\left(\frac{y}{y_c}\right)^2\right],
\end{equation}
and the Prandtl-Schlichting profile \cite{10.7551/mitpress/6781.001.0001,Pope_2000}
\begin{equation}\label{eq:PS}
    \frac{\bar u_x}{\bar{u}_x^m} = \left(1-\left(\frac{y}{C_1y_c}\right)^{3/2}\right)^2,
\end{equation} 
and we see that they capture well the numerical data.

\begin{figure}[htb]
\centering
        \includegraphics[width=0.6\textwidth]{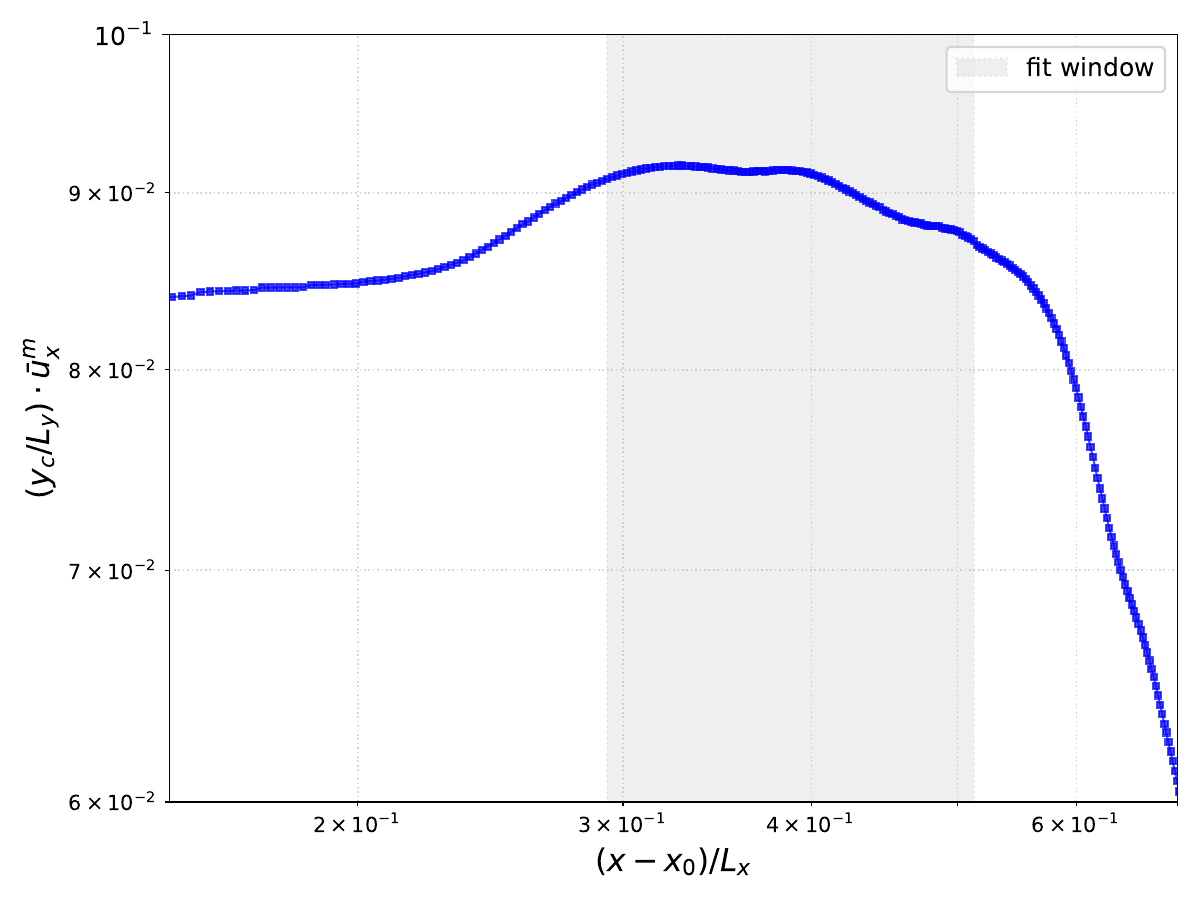}
        \caption{The self-similarity constraint, $\bar{u}_x^m \cdot y_c=const$, is expected to be true in the far-wake regime. The shaded region denotes the streamwise interval used for the similarity analysis. The product $\bar{u}_x^m\cdot y_c$ exhibits an approximately constant plateau over this interval.}
    \label{fig:jecco_wake_constraints}
\end{figure}

As a consistency check, in Fig.~\ref{fig:jecco_wake_constraints} we examine the product $y_c\cdot \bar{u}_x^m$. Classical wake theory predicts that this quantity should be constant in the far-field region~\cite{10.7551/mitpress/6781.001.0001}. Given the periodic boundary conditions in our setup, exact conservation is not expected. Nevertheless, focusing on the far-field region sufficiently distant from both the source and the periodic boundary, the behaviour is broadly consistent with approximate conservation, providing qualitative support for the self-similar interpretation. This is further corroborated by the large-$D$ analysis presented in Appendix~\ref{app:wake_largeD}, where we repeat the same self-similarity analysis for the setup of \cite{Andrade:2019rpn}. In that case, the self-similar fits are noticeably better and the exponent extracted is closer to the classical wake prediction of $1/2$--- this is consistent with the expectation that the larger domain available in the large-$D$ simulation reduces the contamination from periodic boundary conditions.
 
Having established that the boundary response exhibits the expected turbulent wake structure, we now turn to the gravitational description of the dynamics. Within the holographic framework, the bulk geometry provides a geometric encoding of the (strongly coupled) dual fluid evolution, and thus carries an imprint of the turbulent wake. To capture this, we focus on three complementary quantities. First, we study the area element of the apparent horizon, which provides a measure of the local gravitational entropy density. Second, we consider the traceless part of the extrinsic curvature of constant-$r$ hypersurfaces, which characterises anisotropic deformations of the geometry and can be viewed as a measure of bulk geometric strain. Finally, we investigate the full radial structure of the metric perturbations through bulk profiles of the off-diagonal metric component $g_{ty}$, allowing us to track how momentum transport and vortex-induced dynamics extend from the boundary into the interior of the spacetime.

\begin{figure}[thbp]
        \includegraphics[width=\linewidth]{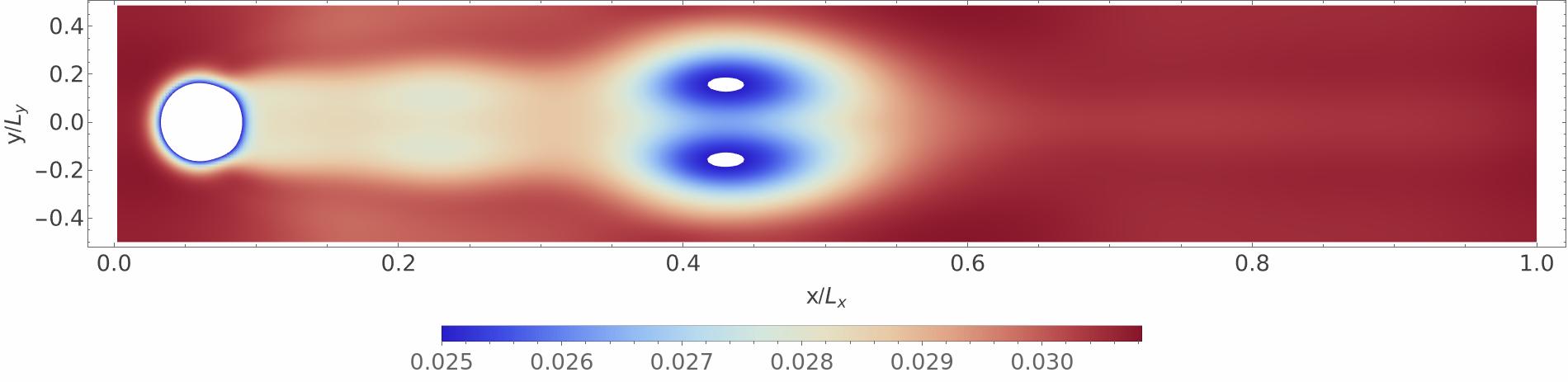}\\
        \includegraphics[width=\linewidth]{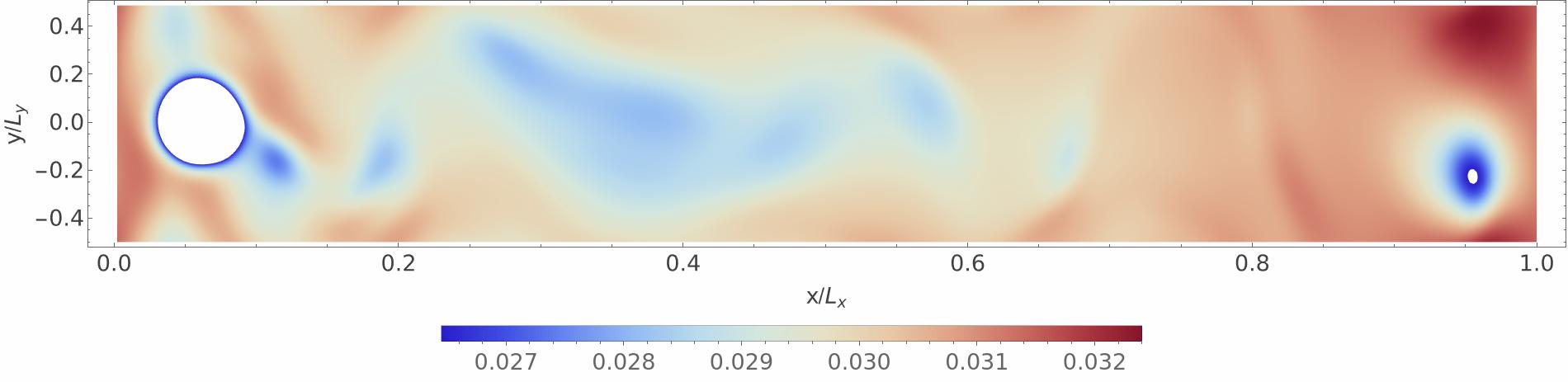}
    \caption{Horizon area element at (Top) $v/v_0=8.672$ and (Bottom) $v/v_0=14.482$. The area element exhibits imprints of the propagation of the vortical structures as well as a strong deformation at the  location corresponding to the source.}
    \label{fig:horizon_area}
\end{figure} 
According to the Bekenstein-Hawking relation, the horizon area is proportional to the gravitational entropy. In equilibrium, this quantity is directly related to the entropy density of the dual field theory. Although the present configuration is driven and therefore out of equilibrium, the horizon area density remains a useful measure of local entropy production and geometric deformation. Fig.~\ref{fig:horizon_area} shows the horizon area density for two snapshots of the evolution, mirroring the vortical features of the boundary flow. This strongly suggests that part of the information associated with the turbulent wake is encoded geometrically through localised deformations of the horizon.

\begin{figure}[htb]    
        \centering
        \includegraphics[width=0.98\linewidth]{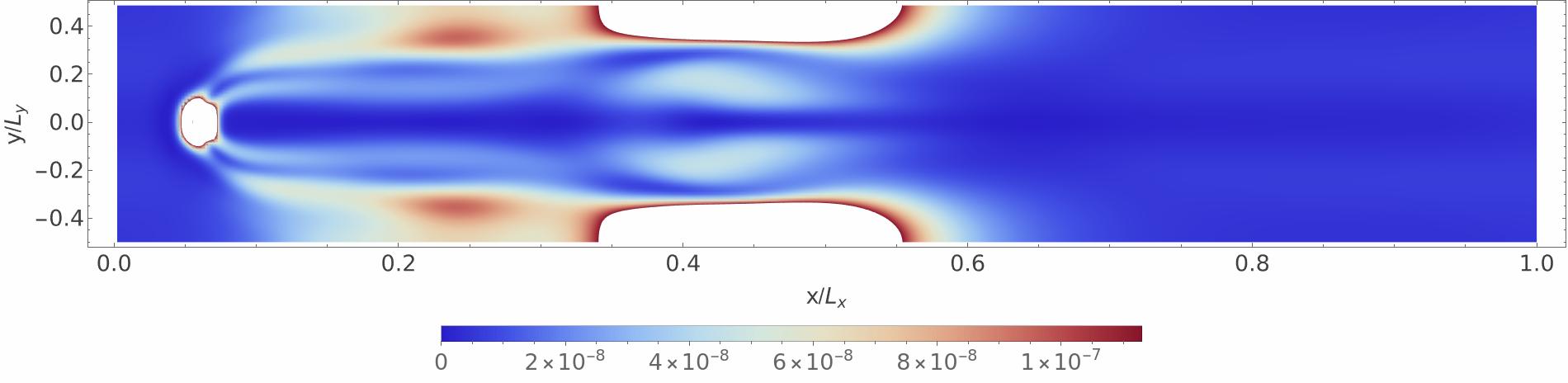}
        \includegraphics[width=0.98\linewidth]{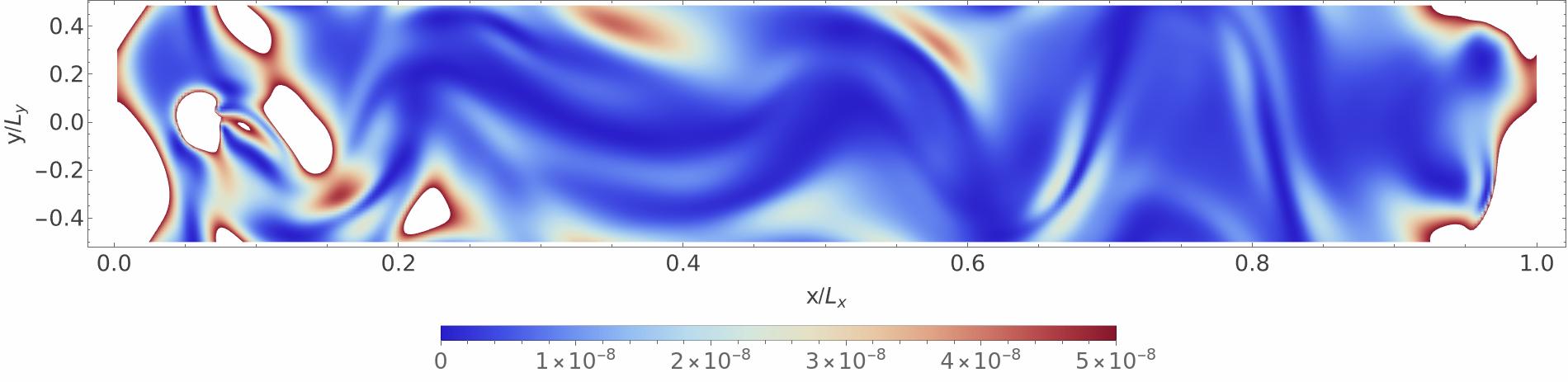}
    \caption{The squared norm of the traceless extrinsic curvature, $\Sigma^2$, at (Top) $v/v_0=8.672$  and (Bottom) $v/v_0=14.482$.}
    \label{fig:horizon_extrinsic_curvature}
\end{figure}

To characterise the geometric response of the horizon beyond its local area density, as done in~\cite{Adams:2013vsa}, we consider the squared norm of the traceless horizon extrinsic curvature,
\begin{align}
\Sigma^2 \equiv \Sigma_{ij}\Sigma^{ij},
\end{align}
where 
\begin{align*}
    & \Sigma^j{}_{i}=K^j{}_{i}-\frac{1}{2}K^k{}_{k}\delta^j{}_i, \qquad
    K_{ij}=P^p{}_iP^q{}_j\nabla_pn_q, \qquad
    P^j{}_i=\delta^j{}_i+l^jn_i, \\
    & n_idx^i=dr, \qquad
    l_idx^i=-dv, \qquad
    l_in^i=-1
\end{align*}
are respectively the traceless extrinsic curvature, the extrinsic curvature, the projector onto the horizon, and the vectors defining the horizon surface. Unlike the horizon area element, which measures the local accumulation of entropy and therefore quantifies isotropic deformations of the horizon, $\Sigma_{ij}$ isolates the anisotropic component. Geometrically, $\Sigma_{ij}$ measures how the null generators of the horizon are stretched and compressed along distinct transverse directions. Consequently, $\Sigma^2$ provides a measure of the local shear of the horizon geometry rather than its overall growth. This is shown in Fig.~\ref{fig:horizon_extrinsic_curvature}. We observe that the resulting profile exhibits a substantially different structure from that of the horizon area density. While the latter is concentrated in the forcing region and in the cores of the largest coherent vortices, the maxima of $\Sigma^2$ are predominantly located along the boundaries of the wake and around the edges of the vortices. This indicates that large $\Sigma^2$ is not associated with regions where the horizon is most displaced from equilibrium, but rather with regions where neighbouring horizon generators undergo substantially different distortions. In the interior of a vortex, the horizon may be significantly, leading to an enhanced area density, while remaining relatively uniform over transverse scales. Such regions therefore contribute only weakly to the traceless extrinsic curvature. In contrast, the interfaces separating the wake from the surrounding flow are characterised by strong spatial gradients and consequently by large anisotropic deformations of the horizon.

From the bulk perspective, $\Sigma^2$ can therefore be interpreted as a measure of the tidal response of the horizon to the forcing. The observed localization of $\Sigma^2$ along the edges of the wake suggests that the gravitational imprint of the turbulent structures is encoded not only in the accumulation of horizon area but also in the shear-induced distortion of the horizon generators. The area density and the traceless extrinsic curvature thus provide complementary probes of the bulk dynamics: the former is sensitive to the magnitude of the horizon deformation, whereas the latter captures its anisotropic structure.

\begin{figure}[tp]
    \centering
    \includegraphics[width=1\linewidth]{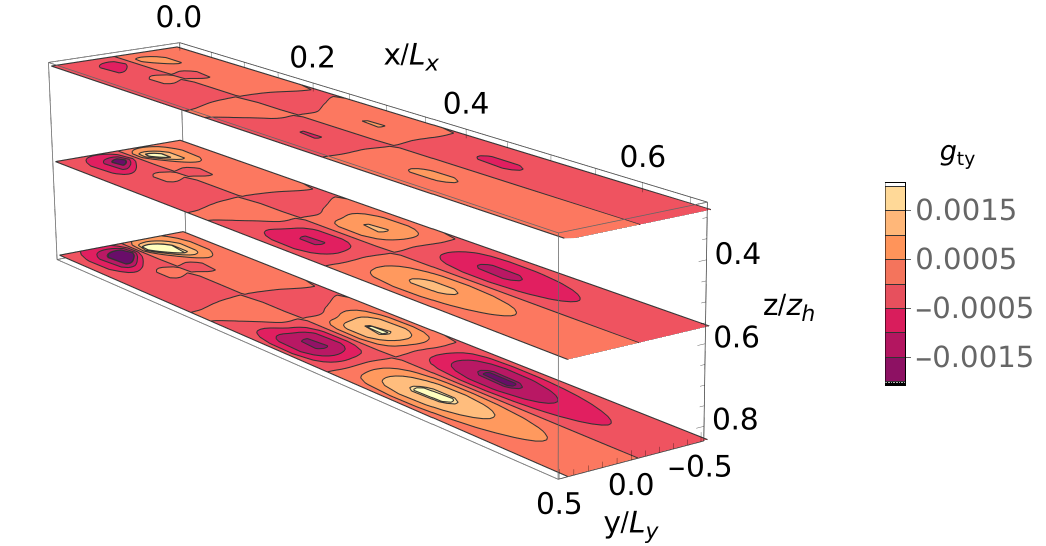}
    \caption{ The $g_{ty}$ component of the metric at different radial slices at $v/v_0=8.672$. The horizon is at $z_h=6$, and $z=0$ represents the AdS boundary. The source is located near the origin~$(0,0)$. We have omitted the portion of the domain that is unperturbed at this time by the wake. The clusters representing vortices extend all the way to the horizon, increasing in size and strength. This indicates that the wake is also imprinted geometrically in the deep interior of the bulk.}
    \label{fig:gtybulk}
\end{figure}

To visualise the propagation of the disturbance through the bulk, we study the metric component $g_{ty}$ at different values of $z=r^{-1}$ in Fig.~\ref{fig:gtybulk}. In the boundary theory this component is associated with momentum transport in the transverse directions and therefore provides a direct probe of the vortical response generated by the source. The three-dimensional bulk profiles reveal how the boundary vortices extend into the radial direction. Rather than being confined to the asymptotic region, the perturbations form coherent structures that penetrate into the interior of the spacetime. The strength as well as the size of these structures increases as one moves deeper into the bulk.

The configuration presented above represents a first exploration of turbulent gravitational wakes, guided by qualitative fluid-dynamical intuition in the absence of a rigorously defined Reynolds number for the holographic system. Beyond the reference simulation discussed above, we have begun exploring other parameter combinations, varying the background velocity $U_x$, the source amplitude $A$, and the characteristic source widths $\sigma_x$, $\sigma_y$;
in some cases we see differences related to the time of detachment of the first pair of vortices and the onset of the $y$-symmetry breaking, but the phenomenology is mostly the same. A systematic study is left for future work. The relevant parameters are: the background velocity $U_x$ and temperature $T$ of the boosted black brane, which together characterise the unperturbed flow; the amplitude $A$, characteristic widths $\sigma_x$, $\sigma_y$, and activation timescales $v_0$ and $\tau$ of the Gaussian source, which determine the strength and geometry of the tidal deformation; and the longitudinal and transverse domain sizes $L_x$ and $L_y$ together with the spatial resolution, which set the range of scales available to the wake. A full exploration of this parameter space would in particular aim to identify the onset of vortex shedding as a function of these parameters, to characterise how the self-similar scaling exponents and the drag force vary across the turbulent regime, and to determine whether an effective Reynolds-like number can be constructed that controls the transition between laminar and turbulent wake dynamics, for instance, a heuristic combination of $U_x$, a characteristic length scale such as $\sigma_x$ or $L_y$, and an effective viscosity estimated from the dual CFT via $\eta/s = 1/4\pi$, which could at least order the scan even in the absence of a rigorously defined holographic Reynolds number. Such a study would also clarify the extent to which the phenomenology reported here is robust to changes in the source geometry and background temperature.

\section{Discussion}\label{sec:disc}

In this work, we studied turbulent gravitational wakes in asymptotically AdS$_4$ spacetimes by solving the full non-linear Einstein equations in the characteristic formulation. Starting from a boosted black brane, we introduced a localized boundary deformation that acts as an obstacle to the flow and generates a gravitational wake. We showed that the resulting boundary dynamics display several characteristic features familiar from turbulent wake phenomenology, including the formation of coherent vortical structures, the subsequent breaking of reflection symmetry, and the development of chaotic downstream dynamics. By constructing suitable time-averaged observables, we further demonstrated that the wake exhibits self-similar behaviour, with the wake width and velocity deficit compatible with the scaling relations expected from classical wake theory. Having established the boundary manifestation of the wake, we then investigated its geometric imprint in the bulk spacetime. In particular, we analysed the horizon area density, the traceless part of the horizon extrinsic curvature, and bulk profiles of metric perturbations, finding that each captures complementary aspects of the wake dynamics. Together, these results provide a detailed picture of how turbulent wake structures are encoded in the geometry of an asymptotically AdS black hole and extend previous studies of gravitational turbulence beyond statistically homogeneous and isotropic configurations. 

A natural quantity to compute is the drag force on the tidal deformation, which provides a direct measure of the rate of $x$-momentum transfer from the source to the wake. Specifically, one considers a contour $\mathcal{C}$ enclosing the source and computes
\begin{equation}
    F_d = \int_{\mathcal{C}} T^{xi} n_i \, dl \, ,
\end{equation}
where $n^i$ is the outward-pointing unit normal to $\mathcal{C}$. When the flow is statistically stationary, conservation of the boundary stress tensor outside the source region implies that $F_d$ is independent of the choice of contour. However, extracting a meaningful contour-independent drag requires the flow between different contours to be statistically stationary, which in turn requires a sufficiently large longitudinal domain so that the wake does not return to the source through the periodic $x$-boundary before stationarity is achieved. This condition is not satisfied in our reference simulation, and we therefore defer a quantitative determination of the drag to future work with larger domains. A full exploration of how $F_d$ depends on background velocity, source amplitude, and obstacle size, and in particular whether it exhibits signatures distinguishing turbulent from laminar wakes, would also be of great interest. 

Another natural direction is to elucidate the relationship between the horizon dynamics observed here and the effective horizon fluid of the membrane paradigm, thereby clarifying how turbulent wake phenomena are represented in the near-horizon description.

From fluid dynamics, we expect the drag to vary qualitatively between turbulent and laminar regimes and thus, the turbulent wake dynamics identified in this work are expected to impact strong-field gravitational processes. In particular, via the AdS/CFT correspondence, these results are relevant to the dynamics of quark-gluon plasmas produced in heavy-ion collisions; for a review on quark-gluon plasma studies within the holographic framework see~\cite{Janik:2010we,Casalderrey-Solana:2011dxg}. In the gravitational context, the tidal deformation sourcing the wake may be viewed as a proxy for a secondary body, suggesting that turbulent wake dynamics could affect the plunge trajectory and associated gravitational waveforms in binary black hole or black hole-neutron star mergers.

Given the direct access we currently have to gravitational waves emitted by black hole mergers and the potential observational ramifications of gravitational turbulence, a natural future direction is to attempt to go beyond AdS asymptotics to explore the possibility of gravitational turbulent wakes in asymptotic flatness. In terms of the equations to be solved, going beyond AdS asymptotics requires removing the cosmological constant from Einstein's equations, which does not change the nature of the partial differential equations that need to be solved. In particular the cosmological constant term is of lower order and thus does not affect local propagation. However, in the absence of a cosmological constant the boundary conditions change too. Black holes in AdS lose energy only through their horizon, while energy in asymptotically flat spacetime can also be lost to infinity. Consequently, in the latter case QNMs decay much faster suggesting a higher effective ``viscosity'' and thus lower  ``Reynolds number'', making turbulence less likely to occur. This, however, is not the case for highly spinning black holes~\cite{Yang:2014tla, Ma:2025rnv} and situations where massive fields could introduce effective boundaries~\cite{Cardoso:2004nk}, making these prime candidates for the study of gravitational turbulence in asymptotically flat spacetimes.  Thus, a natural setup for studying gravitational wakes in asymptotic flatness involves a highly spinning Kerr black hole, where frame dragging plays the role of the background velocity induced by the boosted black brane considered here and the driving is provided by a secondary black hole spiralling around the Kerr solution. Note that for highly spinning Kerr, maintaining the high spin of the primary restricts the setup to an EMRI configuration. From the perspective of the secondary, the Kerr horizon will appear locally flat, even though it is topologically compact.

The behaviour of scalar QNMs in rapidly spinning black holes revealed a phenomenon reminiscent of the inverse cascade displayed in 2+1 dimensional fluids due to resonances. In particular, according to \cite{Yang:2014tla}, when the turbulent instability kicks in, second order modes will increase parametrically and dominate over the leading mode. This is referred to in the literature as parametric instability. Aiming to understand the leading order nonlinear gravitational wave interactions around arbitrarily rapidly rotating Kerr black holes, \cite{Loutrel:2020wbw,Ripley:2020xby} pushed the framework for studying second order perturbations of Kerr black holes in flat spacetime. It would be interesting to take advantage of this framework to explore the parametric instability for gravitational QNM perturbations.

\acknowledgments
It is a pleasure to thank Raimon Luna for collaboration at early stages of this project, and Mikel Sanchez-Garitaonandia for discussions. We would also like to thank Benjamin Withers for comments on the manuscript. G.T. is supported by a UCD Research, Impact and Innovation PhD studentship.
C.P. acknowledges support from a Royal Society -- Research Ireland University Research Fellowship via grant URF/R1/211027. 
M.Z. acknowledges financial support by the Center for Research and Development in Mathematics and Applications (CIDMA) (\url{https://ror.org/05pm2mw36}) through
Fundaç\~ao para a Ci\^encia e a Tecnologia (FCT) (\url{https://ror.org/00snfqn58}), Grants UID/04106/2025 (\url{https://doi.org/10.54499/UID/04106/2025}) and UID/PRR/04106/2025 (\url{https://doi.org/10.54499/UID/PRR/04106/2025}), as well as the projects: Horizon Europe staff exchange (SE) programme HORIZON-MSCA2021-SE-01 Grant No.\ NewFunFiCO-101086251 and 2022.04560.PTDC (\url{https://doi.org/10.54499/2022.04560.PTDC}).
The authors thankfully acknowledge computational resources provided via FCT through projects 2025.09498.CPCA.A3 and 2024.07872.CPCA.A2 (\url{https://doi.org/10.54499/2024.07872.CPCA.A2}) at Deucalion supercomputer, jointly funded by EuroHPC JU and Portugal.

\appendix
\section{Apparent horizon fixing}\label{app:AH}

The apparent horizon (AH) is defined as the hypersurface $r=H(v,x,y)$, where the expansion of outgoing null rays vanishes. The tangent vector to such rays, $l$, is obtained by 
\begin{equation}
    l^\mu=\sqrt{2}s^\mu+n^\mu
\end{equation}
using the ingoing null rays, $n$, together with the form perpendicular to the AH, $s$
\begin{align}
    s_\mu&=-N_s\partial_\mu(H(v,x,y)-r)\,,\\
    n_\mu&=-N_n\partial_r\,.
\end{align}
Note that the normalisation factors $N_s,N_n$ are obtained by demanding  $l^2=0, l\cdot n=-1$ and $s^2=1, s\cdot n=-1/\sqrt{2}$. The expansion is then defined as
\begin{equation}
    \Theta=h^{\mu \nu}\nabla_\mu l_\nu\,.
\end{equation}
where 
\begin{align}
    h_{\mu\nu}=g_{\mu\nu}+l_\mu n_\nu+l_\nu n_\mu\,,
\end{align}
is the induced metric on the hypersurface perpendicular to both ingoing and outgoing rays.  

Demanding that $\Theta=0$ on the hypersurface we arrive to the following expression
\begin{small}
\begin{align}\label{eq:Theta}
    &\Theta=-\Sigma (-H ^{(0,1)} (2 e^B G' H ^{(1,0)} \cosh (\theta)+\cosh (\theta) (2 e^B F_1 G'+e^B \tilde{\theta}+\hat{B}+2 B'
   F_2-F_2')\nonumber\\
   & +\sinh (\theta) (e^B F_1'-2 F_2 G'-\hat{\theta}))+2 e^{2 B} B' F_1 H ^{(1,0)} \cosh (\theta)+e^{2 B}B' (H ^{(1,0)})^2 \cosh (\theta)\nonumber\\
   &+e^{2 B} \tilde{B} H ^{(1,0)} \cosh (\theta)+2 e^{2 B} F_1 G' H ^{(1,0)} \sinh (\theta)-2 e^B F_2
   G' H ^{(1,0)} \cosh (\theta)-e^B F_2' H ^{(1,0)} \sinh (\theta)\nonumber\\
   &+e^{2 B} F_1' H ^{(1,0)} \cosh (\theta)-e^B \hat{\theta} H ^{(1,0)} \cosh (\theta)+e^{2
   B} G' (H ^{(1,0)})^2 \sinh (\theta)+e^{2 B} \tilde{\theta} H ^{(1,0)} \sinh (\theta)\nonumber\\
   &-2 e^B H ^{(1,1)} \sinh (\theta)+e^{2 B} H ^{(2,0)} \cosh
   (\theta)+(H ^{(0,1)})^2 (G' \sinh (\theta)-B' \cosh (\theta))\nonumber\\
   &+H ^{(0,2)} \cosh (\theta)+e^{2 B} B' F_1^2 \cosh (\theta)+e^{2 B} \tilde{B} F_1 \cosh
   (\theta)+e^{2 B} F_1^2 G' \sinh (\theta)\nonumber\\
   &-2 e^B F_1 F_2 G' \cosh (\theta)-e^B F_1 F_2' \sinh (\theta)+e^{2 B} F_1 F_1' \cosh (\theta)-e^B F_1 \hat{\theta} \cosh (\theta)\nonumber\\
   &+e^{2 B} F_1
   \tilde{\theta} \sinh (\theta)-e^B F_2 F_1' \sinh (\theta)-e^B F_2 \tilde{\theta} \cosh (\theta)-e^B \tilde{F_2} \sinh (\theta)-e^B \hat{F_1} \sinh (\theta)\nonumber\\
   &+e^{2 B} \tilde{F_1} \cosh
   (\theta)-\hat{B} F_2 \cosh (\theta)-B' F_2^2 \cosh (\theta)+F_2^2 G' \sinh (\theta)+F_2 F_2' \cosh (\theta)\nonumber\\
   &+F_2 \hat{\theta} \sinh(\theta)+\hat{F_2} \cosh (\theta)) +\Sigma' (2 e^B H ^{(1,0)} (e^B F_1 \cosh (\theta)-F_2 \sinh (\theta))\nonumber\\
   &+2 H ^{(0,1)} (-e^B H ^{(1,0)}
   \sinh (\theta)-e^B F_1 \sinh (\theta)+F_2 \cosh (\theta))+e^{2 B} (H ^{(1,0)})^2 \cosh (\theta)\nonumber\\
   &+(H ^{(0,1)})^2 \cosh (\theta)+e^{2 B} F_1^2 \cosh (\theta)-2
   e^B F_1 F_2 \sinh (\theta)+F_2^2 \cosh (\theta))+2 e^B \Sigma^2 \dot{\Sigma}=0\,.
\end{align}
\end{small}
In this expression we have use the notation
\begin{align*}
    f'&=\partial_r f\\
\dot f&=(\partial_v+\frac{A}{2})f\\
    \hat f&=(\partial_y-F_y \partial_r)f\\
    \tilde f&=( \partial_x- F_x \partial_r)f\,,
\end{align*}
while $H^{(1,0)}\equiv\partial_x H$ and $H^{(0,1)}\equiv\partial_y H$. Following the integration strategy outlined in Section~\ref{subsec:evolution_algorithm}, on the first time slice ($v=0$), given the initial data, we integrate all the remaining metric functions on that slice. We then find the location of the AH;
this is done by linearising equation~\eqref{eq:Theta} around an initial guess and solving iteratively using the Newton-Kantorovich method. Specifically, equation~\eqref{eq:Theta} can be rearranged as
\begin{align}\label{eq:A6}
    &\mathcal{L}(H,\partial H,\partial^2H)=\alpha_{xx}\partial_{xx}H+\alpha_{xy}\partial_{xy}H+\alpha_{yy}\partial_{yy}H\nonumber+\beta_{xx}(\partial_xH)^2\nonumber\\
    &+\beta_{xy}\partial_xH\partial_y H+\beta_{yy}(\partial_{y}H)^2+\gamma_x\partial_xH+\gamma_y\partial_yH+\delta=0\,,
\end{align}
where $\alpha_{xx},\alpha_{xy},\alpha_{yy},\beta_{xx},\beta_{xy},\beta_{yy},\gamma_x,\gamma_y, \delta$ are fully determined in terms of the metric functions (the explicit expressions are omitted for brevity). 
Linearising around a guess solution $H_0(x,y)$, gives
\begin{align}
    &\left(\frac{\partial \mathcal{L}}{\partial H}+\frac{\partial \mathcal{L}}{\partial (\partial_xH)}\partial_x+\frac{\partial \mathcal{L}}{\partial (\partial_yH)}\partial_y+\frac{\partial \mathcal{L}}{\partial (\partial_{xx}H)}\partial_{xx}+\frac{\partial \mathcal{L}}{\partial (\partial_{yy}H)}\partial_{yy}+\frac{\partial \mathcal{L}}{\partial (\partial_{xy}H)}\partial_{xy}\right)_{H=H_0}\delta H\nonumber\\
    &+\mathcal{L}(H_0,\partial H_0, \partial{}^2 H_0)+\mathcal{O}(\delta H^2)=0
\end{align}
with $\delta H(x,y)=H(x,y)-H_0(x,y)$. Computing the derivatives, we obtain a linear equation for $\delta H(x,y)$
\begin{align}\label{eq:C7}
    &(\alpha_{xx}(H_0)\partial_{xx}+\alpha_{xy}(H_0)\partial_{xy}+\alpha_{yy}(H_0)\partial_{yy}+(\gamma_x(H_0)+2\beta_{xx}(H_0)\partial_xH_0+\beta_{xy}(H_0)\partial_yH_0)\partial_x\nonumber\\
    &+(\gamma_y(H_0)+2\beta_{yy}(H_0)\partial_yH_0+\beta_{xy}(H_0)\partial_xH_0)\partial_y+\partial_H\mathcal{L}(H_0))\delta H=-\mathcal{L}(H_0,\partial H_0,\partial^2 H_0)\,.
\end{align}
Solving this equation iteratively until equation~\eqref{eq:A6} is satisfied to a satisfactory precision gives us the location of the AH. 
With $H(x,y)$ at hand, at $v=0$ we adjust the solution on the initial time slice 
\begin{align*}
    \xi&=\xi_\mathrm{init}+\delta \xi\,,\\
    z_h&=\delta \xi+H(x,y)
\end{align*}
so that $\Theta|_{z=z_h}=0$. For $v>0$,  we impose the diffusion-like equation~\eqref{eq:evolvgauge} to ensure that the AH remains at $z=z_h$,
which is recast as an elliptic equation for $\partial_v\xi$
taking the form of equation~\eqref{eq:gauge}.

\section{Tests}
\label{app:tests}

\subsection{Convergence test}
To validate the numerical implementation and assess the reliability of our results, we perform a series of convergence tests following the strategy outlined in Appendix A.3 of~\cite{Bea:2022mfb}. The goal of these tests is to verify that our solutions converge to a well-defined continuum limit as the numerical resolution is increased, and that the observables of interest are insensitive to discretisation artifacts within the parameter ranges explored.

For our present case, since the spatial discretization is performed with a mixture of finite-difference and pseudo-spectral techniques, we fix the number of grid points along the spectral direction and vary only the number of grid points in the uniform grid along the transverse directions $x,y$. The finite-difference operators dominate the numerical error, so the expected convergence rate is controlled by the rate at which we increase the resolution in the uniform grid, as well as the approximation order of the operators. Let us denote by $f$ the solution to the continuum PDE problem and by $f_h$ its numerical approximation $f=f_h+\mathcal{O}(h^n)$, where $h$ is the grid spacing and $n$ the approximation order of the finite-difference operators.

Performing numerical evolutions with coarse, medium and fine resolutions (with grid spacings $h_c, h_m$ and $h_f$ respectively), one can construct the quantity
\begin{align}
    Q=\frac{f_{h_c}-f_{h_m}}{f_{h_m}-f_{h_f}}
\end{align}
often called the convergence factor, which informs us about the rate at which the numerical error induced by the finite-difference scheme converges to zero. Comparison of grid functions corresponding to different resolutions is to be understood by the use of the common grid points among the different resolutions.

Using a physical setup with known exact solution provides a clear benchmark to compare with, and we can prepare such a setup by evolving a homogeneous black brane with only gauge dynamics and vanishing source, $\sigma_0=0$. This can be achieved by using a  different choice for the evolution of the gauge function $\xi$ than the one specified in Section~\ref{subsec:gauge}. In particular, we impose the advection equation
\begin{align}\label{eq:adv}
    \partial_v\xi(v,x,y)=-v_{x}\partial_{x}\xi(v,x,y)
\end{align}
which introduces non-trivial dynamics to the numerical evolution. The only non-vanishing initial data for this setup is the boundary function $a_3$ ($B=\theta=f_x=f_y=0$), which we set to $a_3(v,x,y) =-1$, and the gauge function $\xi$, which we initialize to
\begin{align}
    \xi(0,x,y)=\xi_0+A_x\sin\left(\frac{2 \pi \,n_x}{L}(x_\mathrm{max}-x)\right)
\end{align}
where $L=L_x=L_y$ and $L=x_\mathrm{max}-x_\mathrm{min}$. For such a configuration, the solution to equation \eqref{eq:adv} is
\begin{align}
    \xi(v,x,y) = \xi_0 + A_x \sin\left(  \frac{2 \pi n_{x} } {L_{x}}
    ( x_\mathrm{max} - x+v_{x} v ) 
    \right)
\end{align}
and the exact solution of the metric function $A$ is given by 
\begin{align}
A= \frac{1}{z^2}+2 \frac{\xi}{z}+\xi^2+a_3\frac{z}{1+z \xi}\,.
\end{align}
The location of the apparent horizon $z_h$ can be extracted from $\xi=a_3^{1/3}-1/z_h$.

For this test we use coarse, medium and fine resolutions corresponding to $N=16,32,64$ points in the $x,y$ directions. The $z$ domain is divided in one inner domain $[0,0.1]$ with 12 points and 3 outer domains with 28 points each, where $z_h=1$. The time integration is done with the third-order accurate Adams-Moulton method, where Bogacki-Shampine 3/2 method is used to calculate starting values. We have performed these tests with second-order accurate (periodic) finite-difference operators, with Kreiss-Oliger dissipation $\eta=0.01$ (in contrast with the rest of the simulations in this paper, where $\eta$ is usually bigger). As in~\cite{Bea:2022mfb}, we checked that simulations with two different values of $\eta$ tend to converge when the resolution is increased. In our simulations we used $A_x=0.1, \xi_0=0, n_x=1, x_\mathrm{max}=5,x_\mathrm{min}=-5$, $v_x=1$.

\begin{figure}[thbp]
\centering
    \includegraphics[width=\textwidth]{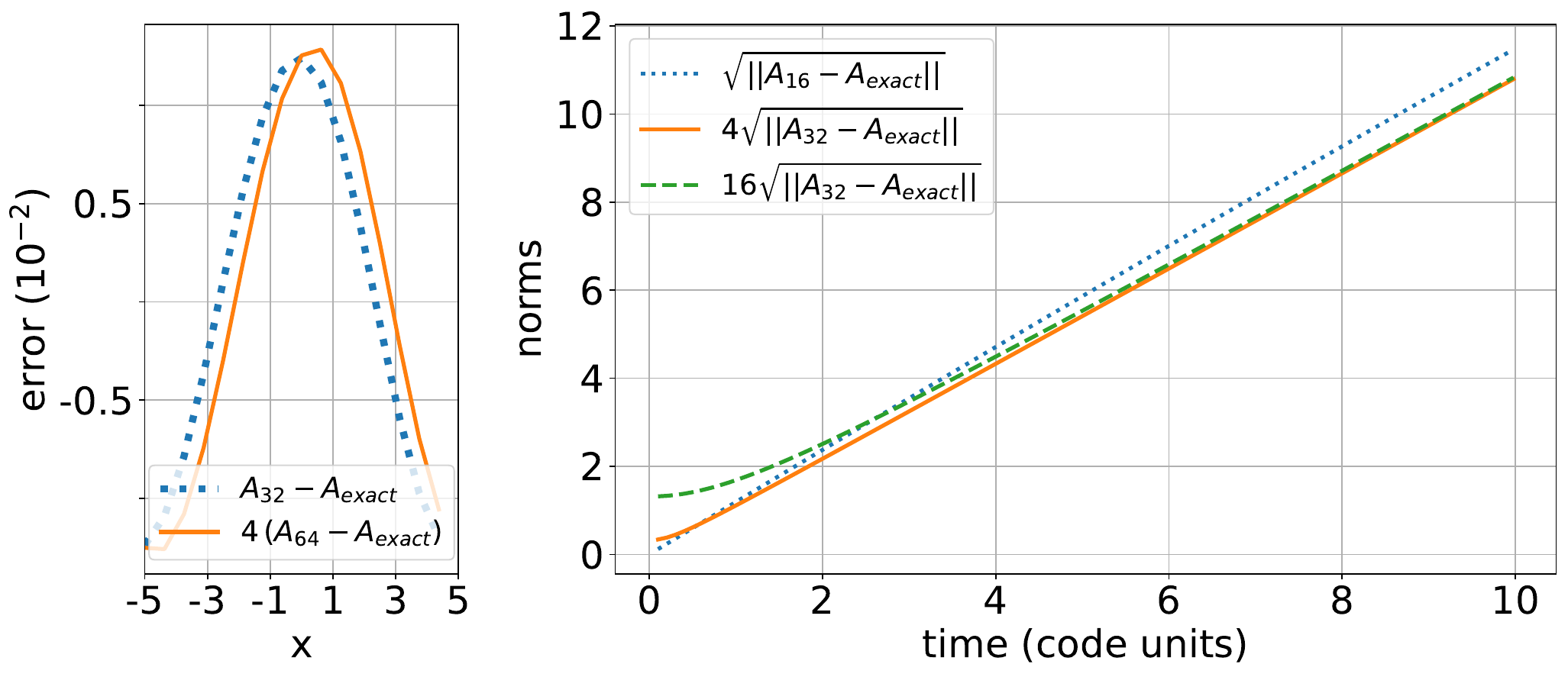}
    \caption{(Left) Pointwise convergence of the metric function $A$ at $v=9.978$, $u=0.8333$ and $y=0.625$. (Right) Norm convergence of the function $A$. We compare the residual between the exact solution and simulations with different resolution. These results are compatible with the expected second-order convergence.
    \label{fig::convergencetest}}
\end{figure}

Convergence tests for the  metric function $A$ can be seen in Fig.~\ref{fig::convergencetest}. As mentioned above, the comparison of the grid functions against the exact solution is performed only on grid points that are common to all three resolutions. The expected convergence factor for this setup is $Q = 4$ for second-order finite difference operators, which is indeed what we observe in the panel on the left. The same convergence rate is expected when we perform a norm comparison. The discretized version of the $L_2$-norm that we employ here is simply the square root of the sum of the squared grid function under consideration (over all domains). In the right panel of the figure we again see very good agreement for the norm convergence rate.

\subsection{Quasinormal Modes}

We now show results for a time-dependent test, where we recover expected QNM frequencies (obtained directly from the equilibrium solution by solving the linear perturbation equations). Specifically, we study two cases, both of which with $L_x=L_y=10, N_x=N_y=64$, with 1 inner and 1 outer domain with $N_z=10$ points each:\\

\noindent
\textbf{Case 1:} We set initial data corresponding to a perturbed homogeneous black brane with vanishing wavenumber, namely $a_3=-1, f_x=f_y=0$, $\xi=0$ on the boundary and $B=0$, $\theta=0.15  z^6$ in the bulk.  The system will relax to the equilibrium state through damped oscillations. In particular, the excited boundary variable $\theta_3$ will evolve in time according to
\begin{equation}
\begin{aligned}
f(v)&=f_\mathrm{eq}+f^{(1)}(v)+\dots\\
f^{(i)}(v)&= A_i\, e^{- \omega_I^{(i)}t} \cos(\omega_R^{(i)}v+\delta_i)\\
\omega^{(i)}&=\omega_R^{(i)}+i\,\omega_I^{(i)}
\end{aligned}
\end{equation}
with the mode with frequency $\omega^{(1)}$ being the longest lived. 
\begin{figure}[htbp]
    \centering
    \includegraphics[width=0.75\linewidth]{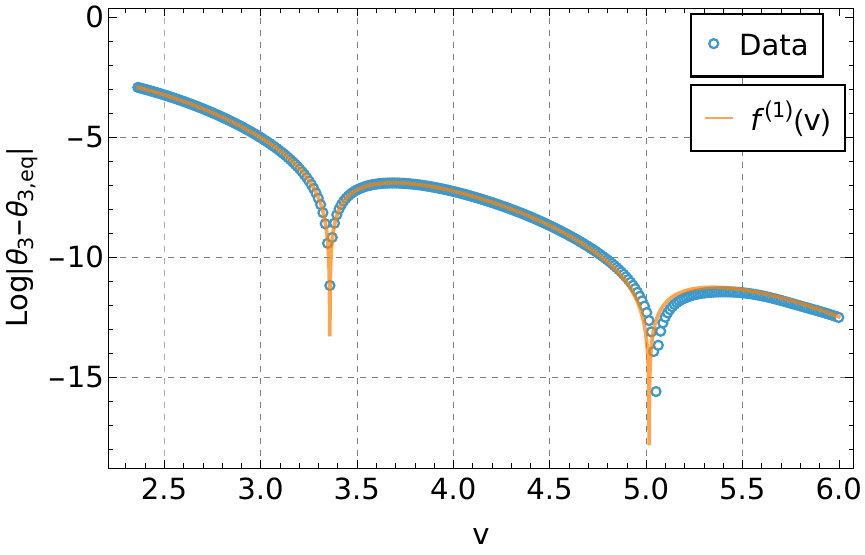}
    \caption{ Time evolution of $\log|\theta_3(v)-\theta_{3, \mathrm{eq}}|$. Fitting the numerical data with a damped sin function, we recover the expected QNM frequency (obtained perturbatively).}
    \label{fig:QNM_tensor}
\end{figure}
In Fig.~\ref{fig:QNM_tensor}, we show a log-plot of $\theta_3-\theta_{3,\mathrm{eq}}$, where $\theta_{3,\mathrm{eq}}$ is the value of $\theta_3$ in equilibrium. At late times, the data clearly behaves as a damped oscillation. We use this fact to fit $f^{(1)}(v)$ to the data.
We find the frequency to be $\omega/(2\pi T)=1.23295-i\,1.7759$, which matches the longest-lived mode (for $k=0, z_h=1$) obtained via shooting method, $\omega/(2\pi T)=1.23263-i\,1.77634$.

\noindent
\textbf{Case 2:} 
We set initial data corresponding to a perturbed homogeneous black brane with non-vanishing wavenumber, corresponding to shear channel perturbations. Without loss of generality we pick $\vec{k}=k_x \hat x$. We initialise the code with $a_3=-1, f_x=0 ,f_y=0.1 \cos(\frac{2\pi}{L_x} x)$, $\xi=0$ and $B=0,\theta=0$.  The system will relax to the equilibrium state through damped oscillations, which we fit following the process described above in order to extract the frequency of the shear mode. Focusing on the boundary variable $f_y $, we find the frequency to be $\omega/(2\pi T) =-i\,0.0908267$, which matches the
longest-lived mode  obtained via shooting, $\omega/(2\pi T)=-i\,0.0908345$. Note that in this case the frequency is purely imaginary and thus the mode is 
purely decaying without oscillations, making the fit straightforward; we 
therefore omit the corresponding plot.

\subsection{Unsourced turbulence}
We have also tested our code in the case of unsourced turbulence, induced by unstable initial data. Specifically, we restrict our simulations to computational domains with $L_x=L_y=L$ and we start  with initial data as in \eqref{eq:match} with velocities given by
\begin{align}
    u_x= A_x \cos\left( \frac{2\pi q_x}{L} y\right) +\delta u_x(\vec x)\,,\quad u_2=0\,,
\end{align}
corresponding to an inhomogeneously boosted black brane. Here $q_x=5$, $A_x=1$, $L=1500$ and $\delta u_x(\vec x)$ is a noise term used to induce the instability given by
\begin{align}
    \delta u_x(\vec x)=\sum_{\vec{m}} B_{\vec{m}} \cos\left(\Delta\phi_{\vec{m}}+\frac{2 \pi (\vec{m}\cdot\vec{x})}{L}\right)
\end{align}
where $\vec{m}, A_{\vec{m}}, \Delta 
\phi_{\vec{m}}$ are chosen from a random sample. In the simulation reported, the sum above includes 2 random modes. The components of the wavenumber $\vec{m}$ were chosen from a uniform distribution of integers running from 1 to 64, while the phase $\Delta\phi_{\vec{m}}$  was chosen from a uniform distribution ranging from 0 to $2\pi$. The amplitude $B_{\vec{m}}$ was chosen from a uniform
distribution ranging from 0 to $B$ with $B=0.1$ a pre-determined parameter.

\begin{figure}[thpb]
    \centering
    \includegraphics[width=0.45\linewidth]{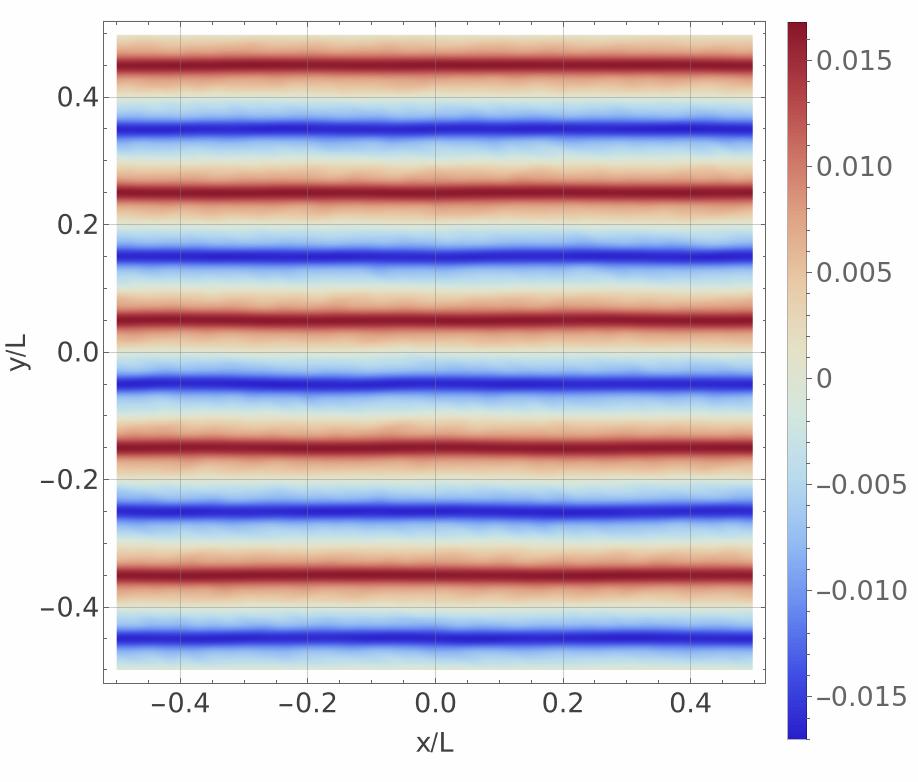}
    \includegraphics[width=0.45\linewidth]{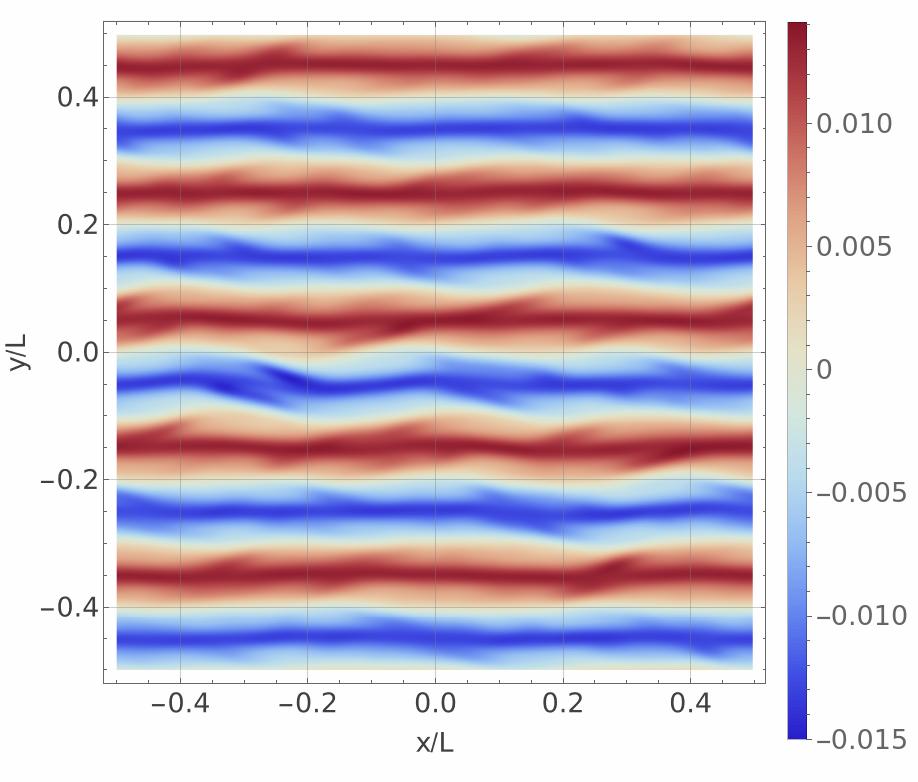}
    \includegraphics[width=0.45\linewidth]{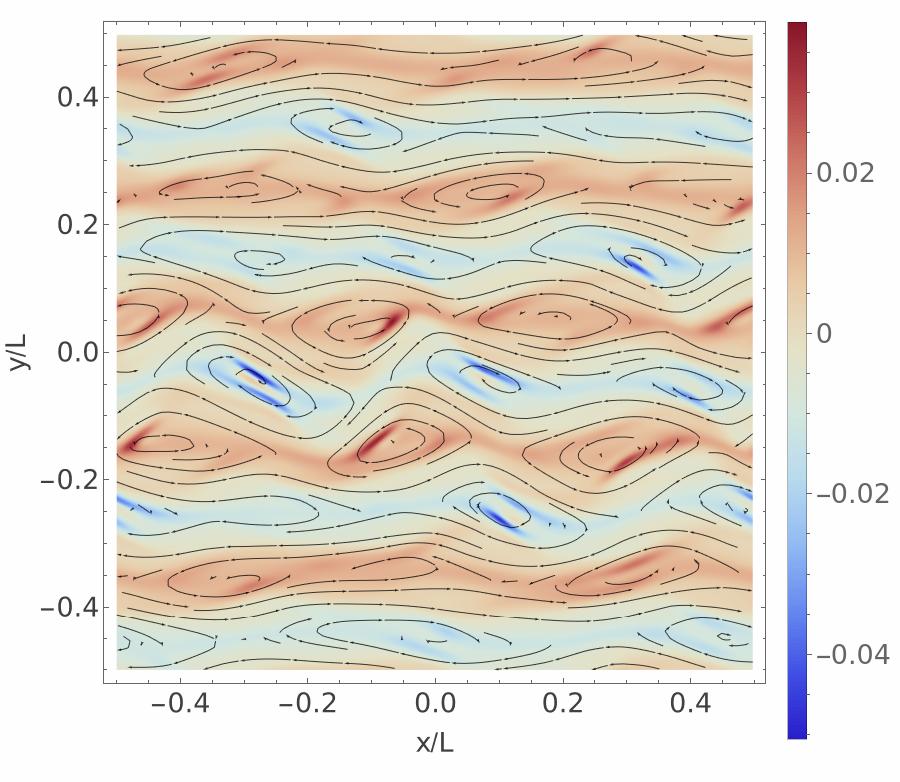}
    \caption{Vorticity of a two dimensional turbulent regime (induced by unstable initial data) at three different times. The instability begins with the formation of vortices at the interface of the counterflow and progressively becomes more widespread. The inverse cascade is apparent in the ``merging" of co-rotating eddies to form larger ones.}
    \label{fig:unsourced_vorticity}
\end{figure}

\begin{figure}[thpb]
    \centering
    \includegraphics[width=0.75\linewidth]{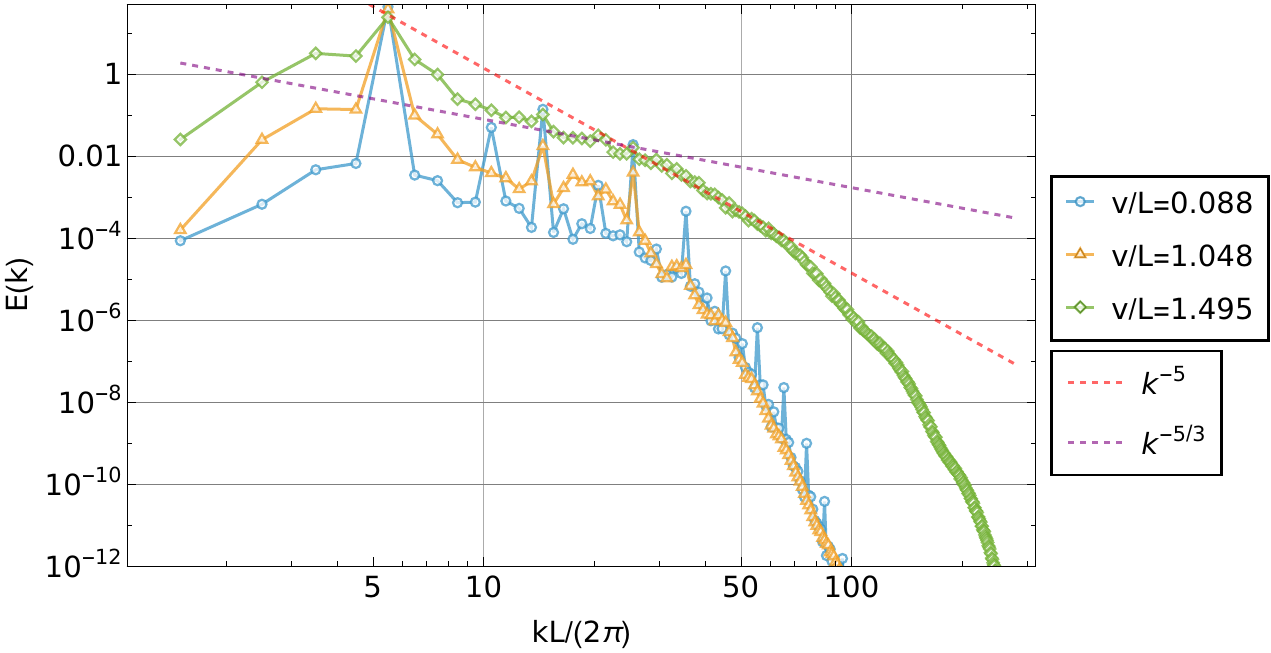}
    \caption{ Kinetic energy power spectrum for the unsourced turbulence simulation shown in Fig.~\ref{fig:unsourced_vorticity} for three different times. ven though there is no explicit homogeneous and isotropic forcing, hints of a power-law cascade are visible within an inertial range that expands toward  smaller $k$ at later times, consistent with the inverse cascade apparent in  the merging of co-rotating eddies seen in Fig.~\ref{fig:unsourced_vorticity}.}
    \label{fig:unsourced_ps}
\end{figure}

Figure~\ref{fig:unsourced_vorticity} shows the vorticity of the corresponding two-dimensional boundary flow at three different times; here $N_x=N_y=300$, while the $z$ domain is divided in one inner domain $[0, 0.1]$
with 12 points and 3 outer domains with 24 points each. The turbulent instability kicks in at around $v/L=0.821$, followed by a chaotic behaviour generating small eddies at $v/L= 1.088$. Then, co-rotating eddies begin to merge with each other, making the inverse cascade apparent. Figure~\ref{fig:unsourced_ps} shows the kinetic energy power spectrum, $E(k)$, computed via 
\begin{equation}
E(k) \equiv \partial_k \int_{|k'|\leq k} \frac{d^dk'}{(2\pi)^d} |\vec{u}_{k'}|^2, 
\end{equation}
 for the same time steps. Even though there is no explicit homogeneous and isotropic forcing in this case, and we thus we do not expect to see a clear inverse cascade compatible with Kolmogorov's $k^{-5/3}$ power law, we do see hints of a power-law cascade within a window  expanding to the smaller $k$ at later times.

\subsection{Turbulent wake: convergence test}

\begin{figure}[h!]
    \centering
    \includegraphics[width=0.45\linewidth]{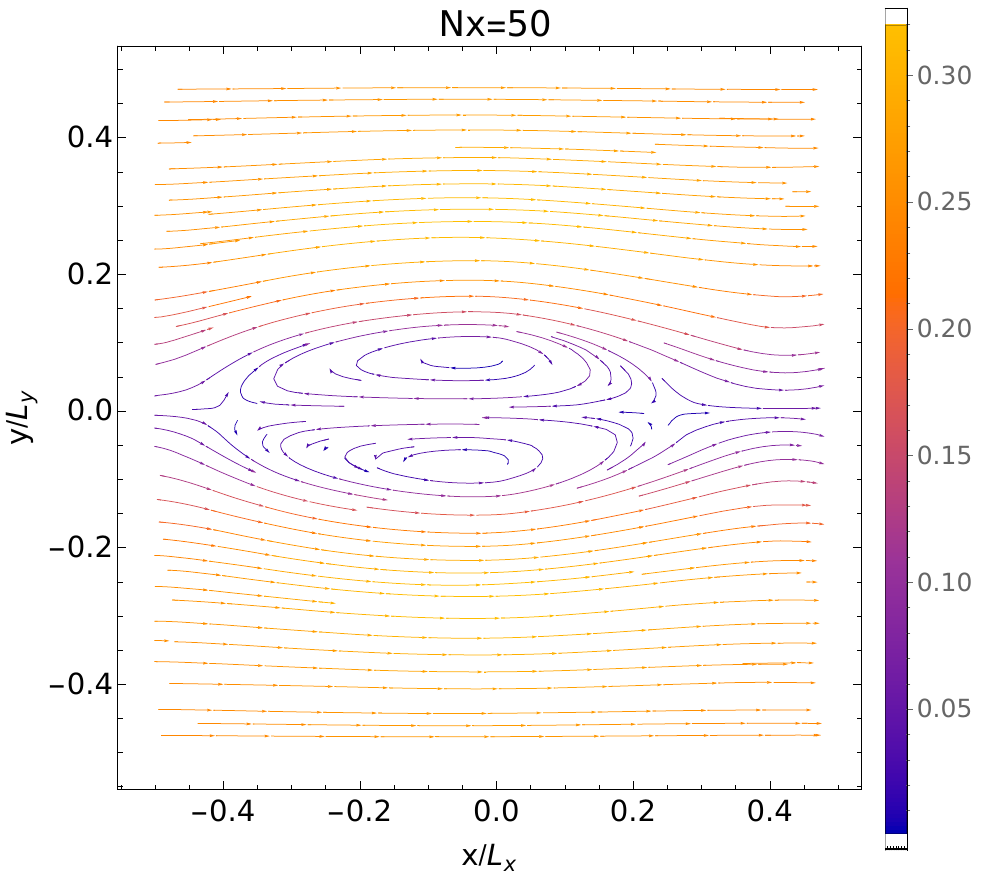}
    \includegraphics[width=0.45\linewidth]{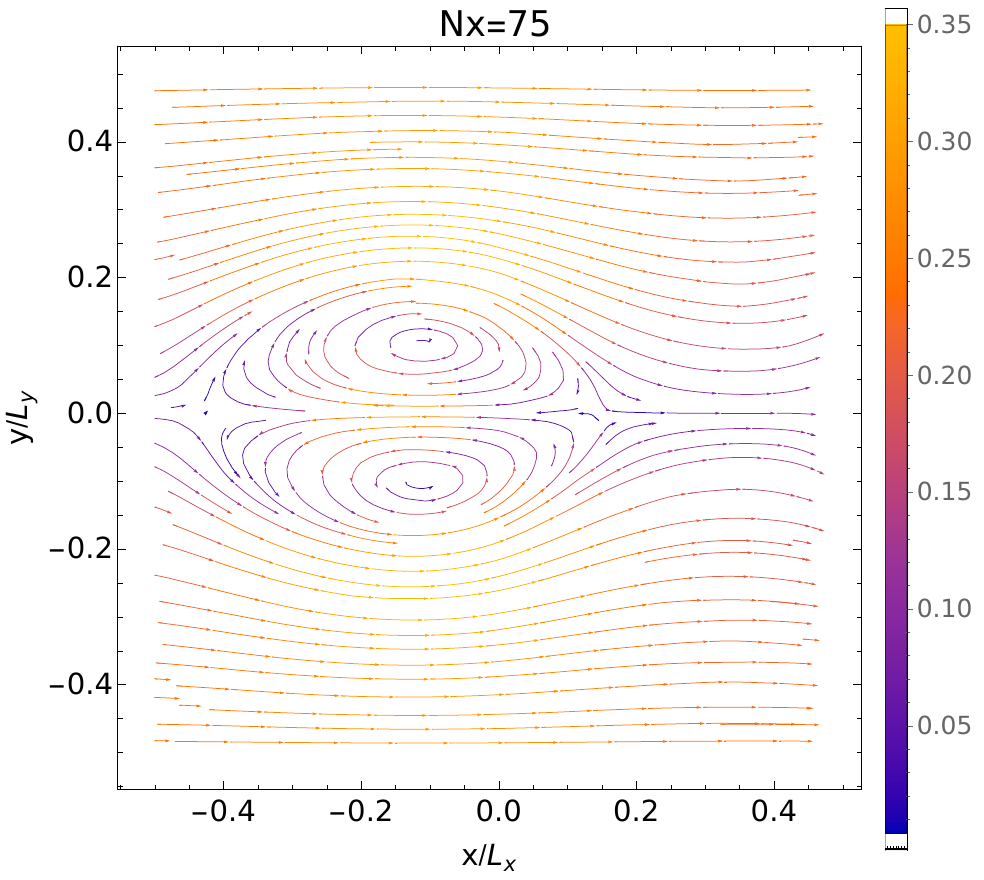}
    \includegraphics[width=0.45\linewidth]{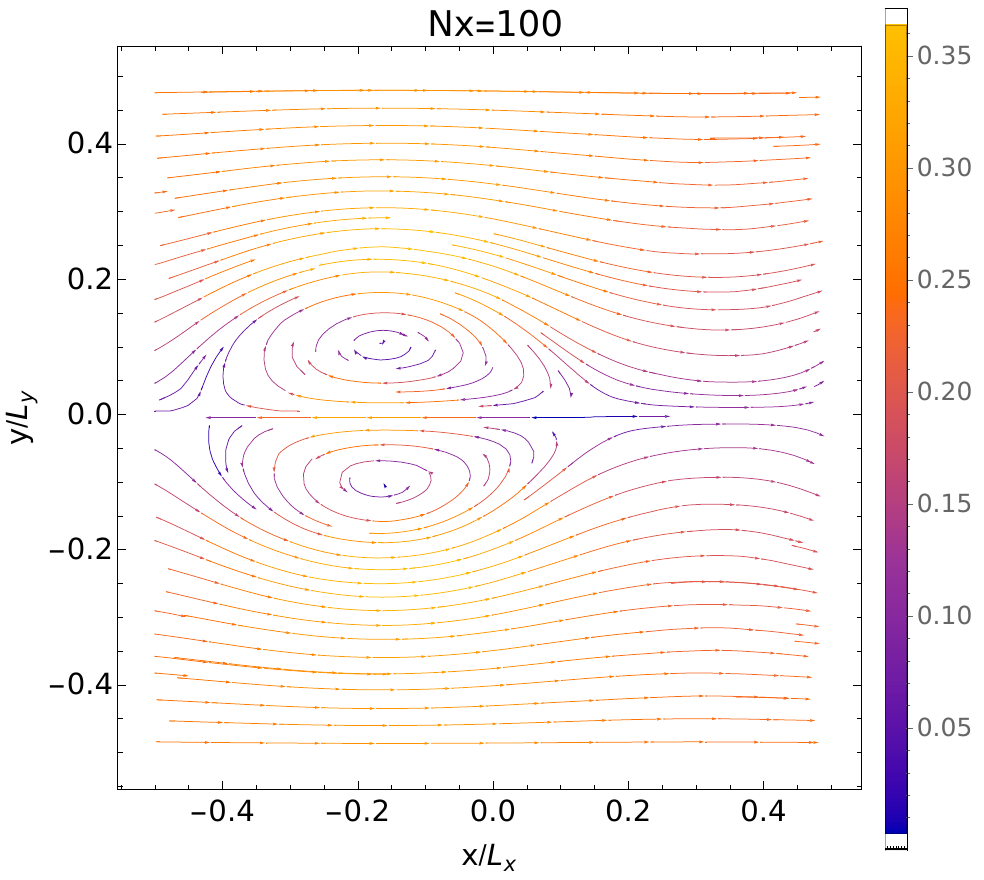}
    \includegraphics[width=0.45\linewidth]{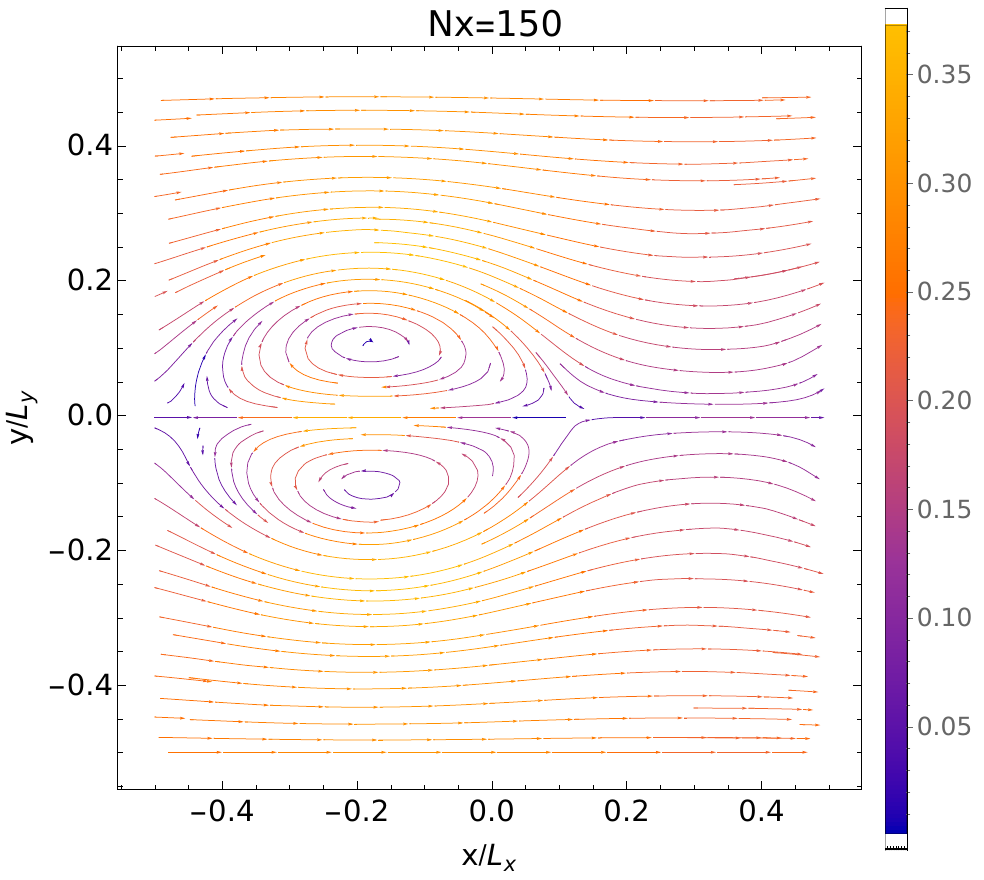}
    \caption{Streamlines of vortices across four different grid resolutions. The figures demonstrate clear convergence as the number of points $N_x$ increases, with the transition from $N_x=50$ to $N_x=75$ points yielding a significantly more resolved and reliable result.}
    \label{fig:wake_convergence}
\end{figure}
Turbulent systems are characterised by dynamics that span a broad hierarchy of spatial and temporal scales, making them intrinsically challenging --- and often computationally prohibitive --- to simulate with uniform numerical resolution. To assess the impact of resolution on our results for the turbulent gravitational wake, we carried out a systematic study for a simulation with the same parameters as in the main text but for a small domain with $L_x=L_y=10^4$, in which the grid spacing was progressively refined. Higher-resolution simulations reveal the formation of vortical structures with larger rotational velocities, see Fig.~\ref{fig:wake_convergence}, reflecting the better capturing of small-scale features. Critically, our results demonstrate that while the lowest resolution is insufficient to capture even qualitative trends, the second-lowest resolution is already sufficient to achieve both qualitative and quantitative convergence. Beyond this threshold, further grid refinement yields diminishing returns, confirming that the overall turbulent wake structure remains robust.

 We note that the convergence study presented above is performed at the level of the qualitative flow structure rather than the fitted self-similar exponent $a$ itself; a dedicated convergence study of the fitted scaling exponents is left for future work, together with the systematic parameter-space exploration discussed in Section 4.

\section{Turbulent wakes in large-$D$}\label{app:wake_largeD}

\begin{figure}[thbp]
     \centering
      \includegraphics[width=1\textwidth]{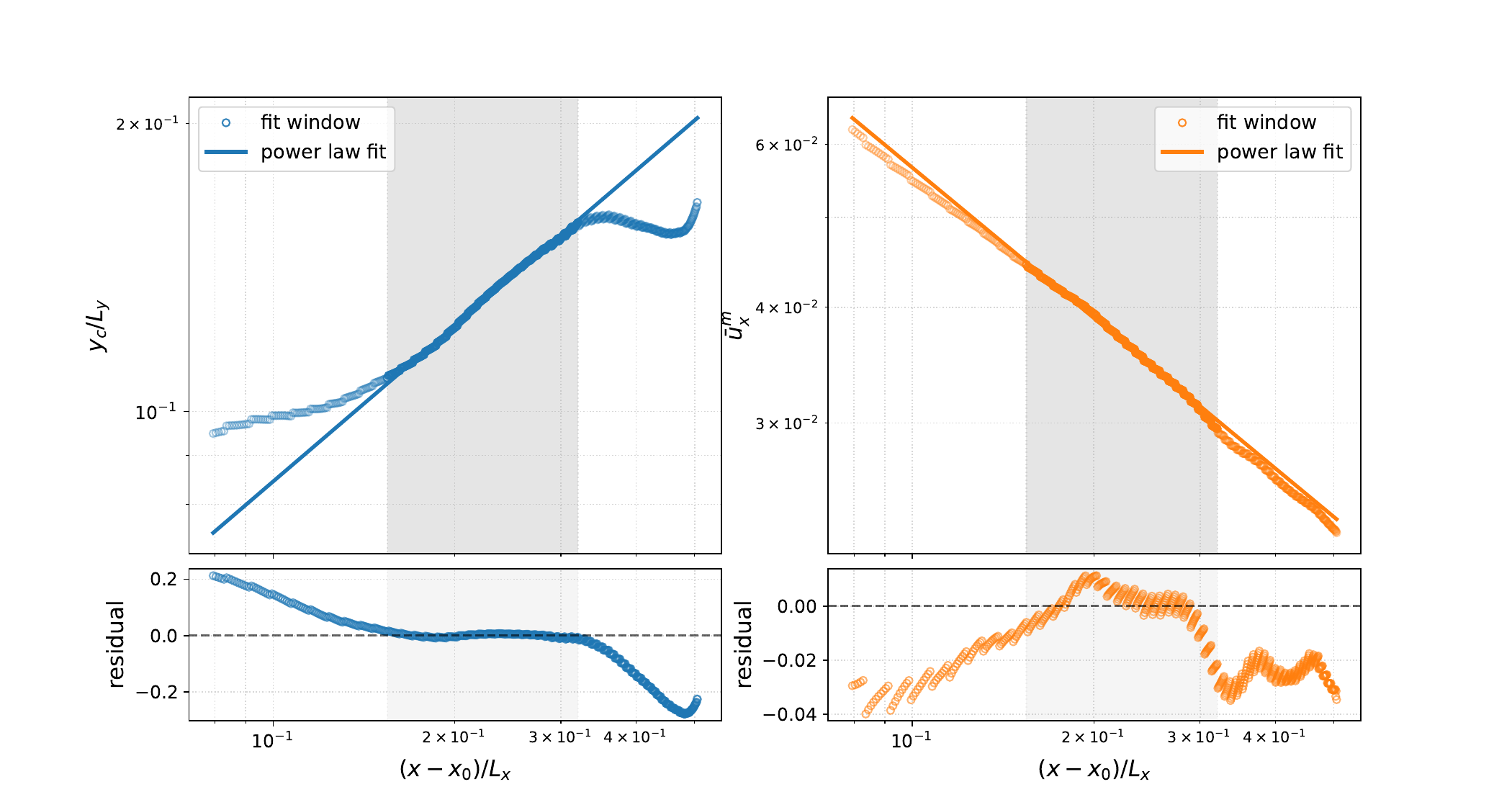}\\
        \includegraphics[width=0.55\textwidth]{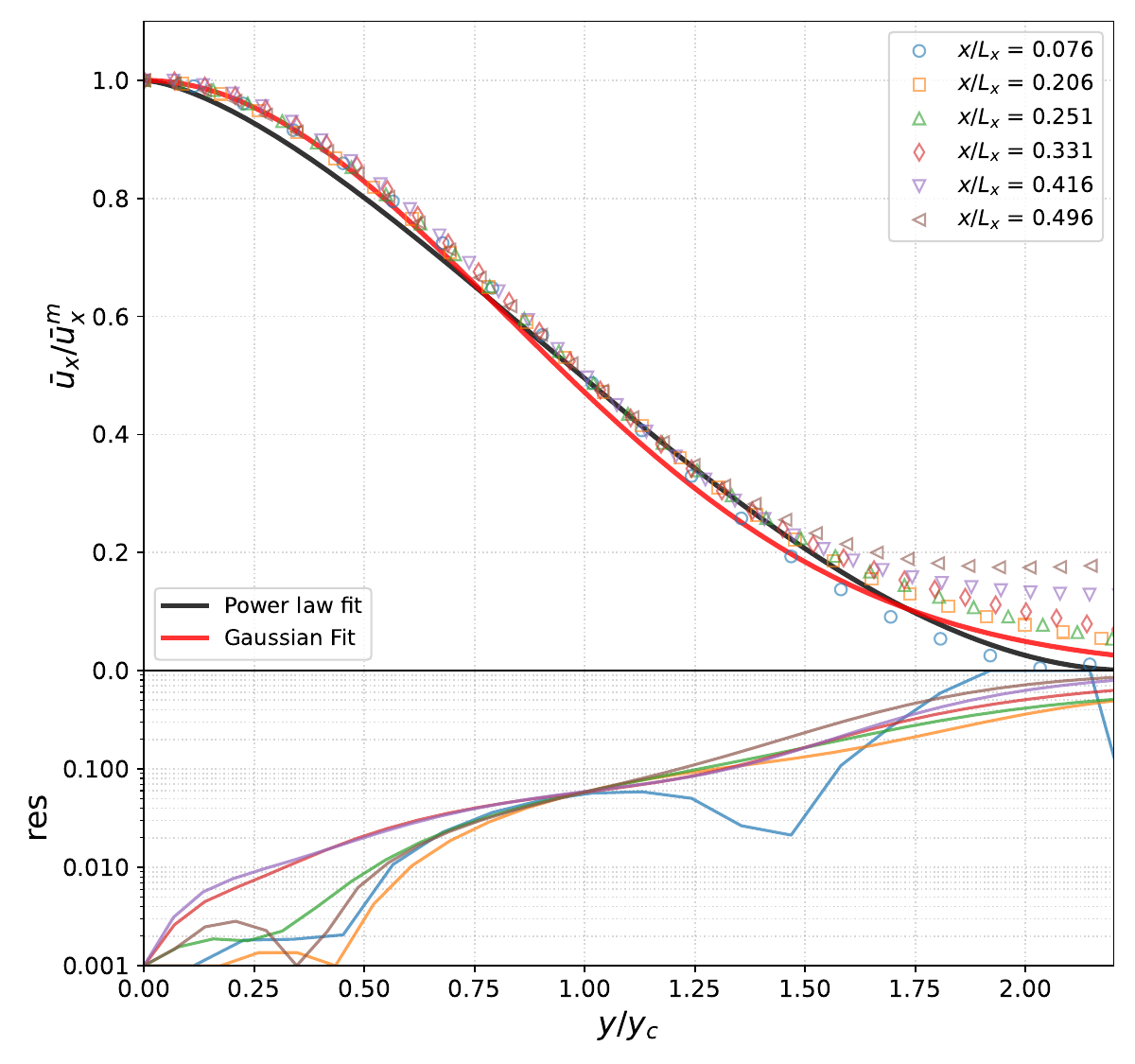}
         \caption{(Left) Logarithmic plot of the wake width, $y_c$, as a function of the $x$, compared with the theoretical prediction. (Right) The velocity deficit, $\bar{u}_x$ as a function of $x$, compared with the theoretical prediction. (Bottom) Self-similar profiles of the wake at distinct $x$ values, where $u/\bar{u}_x^m$ is plotted against $y/y_c$.  We plot only the positive $x$-semi axis since the profiles are mostly symmetric. (All) The lower panels show the relative error between fit and analytical form. The analysis is based on time averages obtained with the methodology described in the main text. }
     \label{fig:SS_largeD}
\end{figure}
Turbulent wakes in the large-$D$ were studied for the first time in \cite{Andrade:2019rpn}. In this appendix, we analyse the self-similar structure of a large-$D$ simulation with $L_x=2\cdot10^5$, $L_y=2\cdot10^4$, $a=1$, $u_x=0.1$, $u_y=0$, $A=-11$, $\sigma_x=\frac{L_y}{8 \pi }=\sigma_y=\frac{L_x}{80 \pi }\sim796$, $v_0=5 \cdot 10^4$, and $\tau=2\cdot 10^4$.

\begin{figure}[h]
     \centering
      \includegraphics[width=0.75\textwidth]{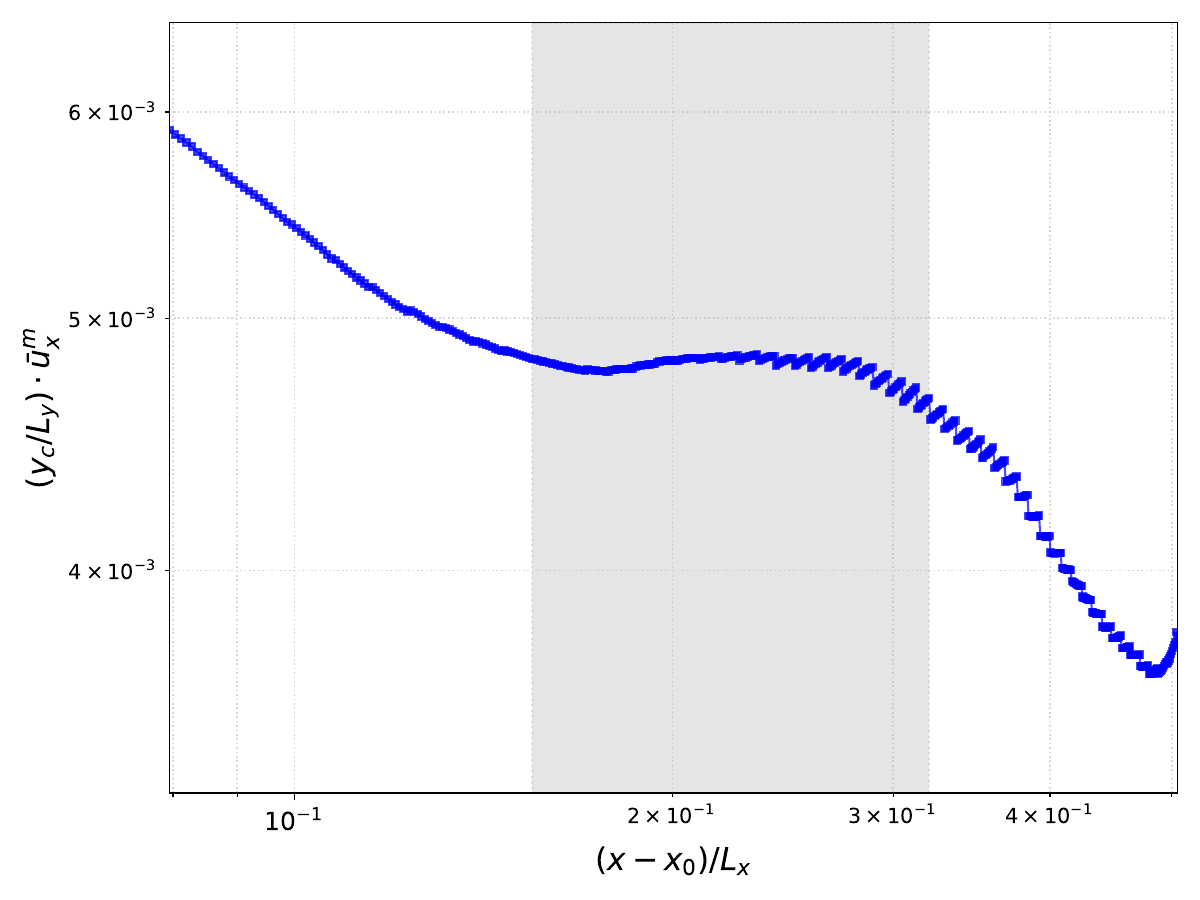}
         \caption{
         The product $y_c\cdot \bar{u}_x^m$          as functions of $x$. The shear free flow condition and the self-similarity assumption require this quantity to be constant in the far field region. }
     \label{fig:constr_largeD}
\end{figure}

In Fig.~\ref{fig:SS_largeD} we plot the velocity deficit and the width of the wake, as well as the self-similar profile. Compared with the four-dimensional simulation presented in the main text, the agreement with the analytical expressions is noticeably better in the large-$D$ case; this is expected given that this simulation has a much larger domain in both the $x$ and $y$ directions. A free power-law fit yields exponents $0.5071\pm 0.001$, a value very close to $1/2$ coming from the classic fluid literature. The constraint, namely the product $y_c\cdot \bar{u}_x^m$, is illustrated in Fig.~\ref{fig:constr_largeD}. The larger domain in the $x$ direction permits a longer averaging interval, thereby improving statistical convergence: during the averaging period, the number of vortices crossing any fixed $x$ location is twice that of the four-dimensional simulations. Taken together, these results support the interpretation that the deviations from perfect self-similarity observed in the main text are primarily a consequence of the limited domain size and averaging window available in the four-dimensional simulation, rather than a failure of the self-similar wake phenomenology itself.

\bibliography{turbs}{}

\providecommand{\href}[2]{#2}\begingroup\raggedright\begin{thebibliography}{10}

\bibitem{K41a}
A.~{Kolmogorov}, \emph{The local structure of turbulence in incompressible
  viscous fluid for very large reynolds' numbers}, {\emph{Akademiia Nauk SSSR
  Doklady} {\bfseries 30} (1941) 301}.

\bibitem{K41b}
A.~N. {Kolmogorov}, \emph{Dissipation of energy in locally isotropic
  turbulence}, {\emph{Akademiia Nauk SSSR Doklady} {\bfseries 32} (1941) 16}.

\bibitem{Maldacena:1997re}
J.~M. Maldacena, \emph{{The Large N limit of superconformal field theories and
  supergravity}}, \href{https://doi.org/10.1023/A:1026654312961,
  10.4310/ATMP.1998.v2.n2.a1}{\emph{Int. J. Theor. Phys.} {\bfseries 38} (1999)
  1113} [\href{https://arxiv.org/abs/hep-th/9711200}{{\ttfamily
  hep-th/9711200}}].

\bibitem{Carrasco:2012nf}
F.~Carrasco, L.~Lehner, R.~C. Myers, O.~Reula and A.~Singh, \emph{{Turbulent
  flows for relativistic conformal fluids in 2+1 dimensions}},
  \href{https://doi.org/10.1103/PhysRevD.86.126006}{\emph{Phys. Rev.}
  {\bfseries D86} (2012) 126006}
  [\href{https://arxiv.org/abs/1210.6702}{{\ttfamily 1210.6702}}].

\bibitem{Green:2013zba}
S.~R. Green, F.~Carrasco and L.~Lehner, \emph{{Holographic Path to the
  Turbulent Side of Gravity}},
  \href{https://doi.org/10.1103/PhysRevX.4.011001}{\emph{Phys. Rev.} {\bfseries
  X4} (2014) 011001} [\href{https://arxiv.org/abs/1309.7940}{{\ttfamily
  1309.7940}}].

\bibitem{Adams:2013vsa}
A.~Adams, P.~M. Chesler and H.~Liu, \emph{{Holographic turbulence}},
  \href{https://doi.org/10.1103/PhysRevLett.112.151602}{\emph{Phys. Rev. Lett.}
  {\bfseries 112} (2014) 151602}
  [\href{https://arxiv.org/abs/1307.7267}{{\ttfamily 1307.7267}}].

\bibitem{Emparan:2015hwa}
R.~Emparan, T.~Shiromizu, R.~Suzuki, K.~Tanabe and T.~Tanaka, \emph{{Effective
  theory of Black Holes in the 1/D expansion}},
  \href{https://doi.org/10.1007/JHEP06(2015)159}{\emph{JHEP} {\bfseries 06}
  (2015) 159} [\href{https://arxiv.org/abs/1504.06489}{{\ttfamily
  1504.06489}}].

\bibitem{Emparan:2015gva}
R.~Emparan, R.~Suzuki and K.~Tanabe, \emph{{Evolution and End Point of the
  Black String Instability: Large D Solution}},
  \href{https://doi.org/10.1103/PhysRevLett.115.091102}{\emph{Phys. Rev. Lett.}
  {\bfseries 115} (2015) 091102}
  [\href{https://arxiv.org/abs/1506.06772}{{\ttfamily 1506.06772}}].

\bibitem{Bhattacharyya:2015dva}
S.~Bhattacharyya, A.~De, S.~Minwalla, R.~Mohan and A.~Saha, \emph{{A membrane
  paradigm at large D}},
  \href{https://doi.org/10.1007/JHEP04(2016)076}{\emph{JHEP} {\bfseries 04}
  (2016) 076} [\href{https://arxiv.org/abs/1504.06613}{{\ttfamily
  1504.06613}}].

\bibitem{Emparan:2016sjk}
R.~Emparan, K.~Izumi, R.~Luna, R.~Suzuki and K.~Tanabe, \emph{{Hydro-elastic
  Complementarity in Black Branes at large D}},
  \href{https://doi.org/10.1007/JHEP06(2016)117}{\emph{JHEP} {\bfseries 06}
  (2016) 117} [\href{https://arxiv.org/abs/1602.05752}{{\ttfamily
  1602.05752}}].

\bibitem{PhysRevD.31.725}
W.~A. Hiscock and L.~Lindblom, \emph{Generic instabilities in first-order
  dissipative relativistic fluid theories},
  \href{https://doi.org/10.1103/PhysRevD.31.725}{\emph{Phys. Rev. D} {\bfseries
  31} (1985) 725}.

\bibitem{Poovuttikul:2019ckt}
N.~Poovuttikul and W.~Sybesma, \emph{{First order non-Lorentzian fluids,
  entropy production and linear instabilities}},
  \href{https://doi.org/10.1103/PhysRevD.102.065007}{\emph{Phys. Rev. D}
  {\bfseries 102} (2020) 065007}
  [\href{https://arxiv.org/abs/1911.00010}{{\ttfamily 1911.00010}}].

\bibitem{Rozali:2017bll}
M.~Rozali, E.~Sabag and A.~Yarom, \emph{{Holographic Turbulence in a Large
  Number of Dimensions}},
  \href{https://doi.org/10.1007/JHEP04(2018)065}{\emph{JHEP} {\bfseries 04}
  (2018) 065} [\href{https://arxiv.org/abs/1707.08973}{{\ttfamily
  1707.08973}}].

\bibitem{Andrade:2019rpn}
T.~Andrade, C.~Pantelidou, J.~Sonner and B.~Withers, \emph{{Driven black holes:
  from Kolmogorov scaling to turbulent wakes}},
  \href{https://doi.org/10.1007/JHEP07(2021)063}{\emph{JHEP} {\bfseries 07}
  (2021) 063} [\href{https://arxiv.org/abs/1912.00032}{{\ttfamily
  1912.00032}}].

\bibitem{Westernacher-Schneider:2015gfa}
J.~R. Westernacher-Schneider, L.~Lehner and Y.~Oz, \emph{{Scaling Relations in
  Two-Dimensional Relativistic Hydrodynamic Turbulence}},
  \href{https://doi.org/10.1007/JHEP12(2015)067}{\emph{JHEP} {\bfseries 12}
  (2015) 067} [\href{https://arxiv.org/abs/1510.00736}{{\ttfamily
  1510.00736}}].

\bibitem{Westernacher-Schneider:2017snn}
J.~R. Westernacher-Schneider and L.~Lehner, \emph{{Numerical Measurements of
  Scaling Relations in Two-Dimensional Conformal Fluid Turbulence}},
  \href{https://doi.org/10.1007/JHEP08(2017)027}{\emph{JHEP} {\bfseries 08}
  (2017) 027} [\href{https://arxiv.org/abs/1706.07480}{{\ttfamily
  1706.07480}}].

\bibitem{Waeber:2021xba}
S.~Waeber and A.~Yarom, \emph{{Stochastic gravity and turbulence}},
  \href{https://doi.org/10.1007/JHEP12(2021)185}{\emph{JHEP} {\bfseries 12}
  (2021) 185} [\href{https://arxiv.org/abs/2105.01551}{{\ttfamily
  2105.01551}}].

\bibitem{Oz:2024smd}
Y.~Oz, S.~Waeber and A.~Yarom, \emph{{Holographic turbulence from a random
  gravitational potential}},
  \href{https://doi.org/10.1007/JHEP08(2024)071}{\emph{JHEP} {\bfseries 08}
  (2024) 071} [\href{https://arxiv.org/abs/2402.08471}{{\ttfamily
  2402.08471}}].

\bibitem{Scott2007Nonrobustness}
R.~K. Scott, \emph{Nonrobustness of the two-dimensional turbulent inverse
  cascade}, \href{https://doi.org/10.1103/PhysRevE.75.046301}{\emph{Physical
  Review E} {\bfseries 75} (2007) 046301}.

\bibitem{Du:2025vgc}
J.~Du, Y.~Tian and H.~Zhang, \emph{{Holographic Turbulence and the Fractal
  Dimension of the Turbulent Horizon}},
  \href{https://arxiv.org/abs/2510.12198}{{\ttfamily 2510.12198}}.

\bibitem{Bea:2022mfb}
Y.~Bea, J.~Casalderrey-Solana, T.~Giannakopoulos, A.~Jansen, D.~Mateos,
  M.~Sanchez-Garitaonandia and M.~Zilh{\~a}o, \emph{{Holographic bubbles with
  Jecco: expanding, collapsing and critical}},
  \href{https://doi.org/10.1007/JHEP09(2022)008}{\emph{JHEP} {\bfseries 09}
  (2022) 008} [\href{https://arxiv.org/abs/2202.10503}{{\ttfamily
  2202.10503}}].

\bibitem{Winicour:2012znc}
J.~Winicour, \emph{{Characteristic Evolution and Matching}},
  \href{https://doi.org/10.12942/lrr-2012-2}{\emph{Living Rev. Rel.} {\bfseries
  15} (2012) 2}.

\bibitem{Chesler:2013lia}
P.~M. Chesler and L.~G. Yaffe, \emph{{Numerical solution of gravitational
  dynamics in asymptotically anti-de Sitter spacetimes}},
  \href{https://doi.org/10.1007/JHEP07(2014)086}{\emph{JHEP} {\bfseries 07}
  (2014) 086} [\href{https://arxiv.org/abs/1309.1439}{{\ttfamily 1309.1439}}].

\bibitem{Bea:2021zol}
Y.~Bea, J.~Casalderrey-Solana, T.~Giannakopoulos, A.~Jansen, S.~Krippendorf,
  D.~Mateos, M.~Sanchez-Garitaonandia and M.~Zilh{\~a}o, \emph{{Spinodal
  Gravitational Waves}},
  \href{https://doi.org/10.1007/JHEP11(2025)093}{\emph{JHEP} {\bfseries 11}
  (2025) 093} [\href{https://arxiv.org/abs/2112.15478}{{\ttfamily
  2112.15478}}].

\bibitem{Bea:2021ieq}
Y.~Bea, J.~Casalderrey-Solana, T.~Giannakopoulos, D.~Mateos,
  M.~Sanchez-Garitaonandia and M.~Zilh{\~a}o, \emph{{Domain collisions}},
  \href{https://doi.org/10.1007/JHEP06(2022)025}{\emph{JHEP} {\bfseries 06}
  (2022) 025} [\href{https://arxiv.org/abs/2111.03355}{{\ttfamily
  2111.03355}}].

\bibitem{Bea:2021zsu}
Y.~Bea, J.~Casalderrey-Solana, T.~Giannakopoulos, D.~Mateos,
  M.~Sanchez-Garitaonandia and M.~Zilh{\~a}o, \emph{{Bubble wall velocity from
  holography}}, \href{https://doi.org/10.1103/PhysRevD.104.L121903}{\emph{Phys.
  Rev. D} {\bfseries 104} (2021) L121903}
  [\href{https://arxiv.org/abs/2104.05708}{{\ttfamily 2104.05708}}].

\bibitem{Bea:2024bls}
Y.~Bea, M.~Giliberti, D.~Mateos, M.~Sanchez-Garitaonandia, A.~Serantes and
  M.~Zilh{\~a}o, \emph{{Bubble dynamics in a QCD-like phase diagram}},
  \href{https://doi.org/10.1007/JHEP04(2026)013}{\emph{JHEP} {\bfseries 04}
  (2026) 013} [\href{https://arxiv.org/abs/2412.09588}{{\ttfamily
  2412.09588}}].

\bibitem{Chesler:2010bi}
P.~M. Chesler and L.~G. Yaffe, \emph{{Holography and colliding gravitational
  shock waves in asymptotically AdS$_{5}$ spacetime}},
  \href{https://doi.org/10.1103/PhysRevLett.106.021601}{\emph{Phys. Rev. Lett.}
  {\bfseries 106} (2011) 021601}
  [\href{https://arxiv.org/abs/1011.3562}{{\ttfamily 1011.3562}}].

\bibitem{Attems:2017zam}
M.~Attems, J.~Casalderrey-Solana, D.~Mateos, D.~Santos-Oliv{\'a}n, C.~F.
  Sopuerta, M.~Triana and M.~Zilh{\~a}o, \emph{{Paths to equilibrium in
  non-conformal collisions}},
  \href{https://doi.org/10.1007/JHEP06(2017)154}{\emph{JHEP} {\bfseries 06}
  (2017) 154} [\href{https://arxiv.org/abs/1703.09681}{{\ttfamily
  1703.09681}}].

\bibitem{Balasubramanian:2013yqa}
K.~Balasubramanian and C.~P. Herzog, \emph{{Losing Forward Momentum
  Holographically}},
  \href{https://doi.org/10.1088/0264-9381/31/12/125010}{\emph{Class. Quant.
  Grav.} {\bfseries 31} (2014) 125010}
  [\href{https://arxiv.org/abs/1312.4953}{{\ttfamily 1312.4953}}].

\bibitem{Balasubramanian_1999}
V.~Balasubramanian and P.~Kraus, \emph{A stress tensor for anti-de sitter
  gravity}, \href{https://doi.org/10.1007/s002200050764}{\emph{Communications
  in Mathematical Physics} {\bfseries 208} (1999) 413–428}.

\bibitem{Andrade:2018zeb}
T.~Andrade, C.~Pantelidou and B.~Withers, \emph{{Large D holography with metric
  deformations}}, \href{https://doi.org/10.1007/JHEP09(2018)138}{\emph{JHEP}
  {\bfseries 09} (2018) 138}
  [\href{https://arxiv.org/abs/1806.00306}{{\ttfamily 1806.00306}}].

\bibitem{10.7551/mitpress/6781.001.0001}
G.~N. Abramovich, \emph{The Theory of Turbulent Jets}. The MIT Press, 03, 2003,
  \href{https://doi.org/10.7551/mitpress/6781.001.0001}{10.7551/mitpress/6781.001.0001}.

\bibitem{Pope_2000}
S.~B. Pope, \emph{Turbulent Flows}. Cambridge University Press, 2000.

\bibitem{Janik:2010we}
R.~A. Janik, \emph{{The Dynamics of Quark-Gluon Plasma and AdS/CFT}},
  \href{https://doi.org/10.1007/978-3-642-04864-7_5}{\emph{Lect. Notes Phys.}
  {\bfseries 828} (2011) 147}
  [\href{https://arxiv.org/abs/1003.3291}{{\ttfamily 1003.3291}}].

\bibitem{Casalderrey-Solana:2011dxg}
J.~Casalderrey-Solana, H.~Liu, D.~Mateos, K.~Rajagopal and U.~Achim~Wiedemann,
  \emph{{Gauge/String Duality, Hot QCD and Heavy Ion Collisions}}. Cambridge
  University Press, 2014,
  \href{https://doi.org/10.1017/9781009403504}{10.1017/9781009403504},
  [\href{https://arxiv.org/abs/1101.0618}{{\ttfamily 1101.0618}}].

\bibitem{Yang:2014tla}
H.~Yang, A.~Zimmerman and L.~Lehner, \emph{{Turbulent Black Holes}},
  \href{https://doi.org/10.1103/PhysRevLett.114.081101}{\emph{Phys. Rev. Lett.}
  {\bfseries 114} (2015) 081101}
  [\href{https://arxiv.org/abs/1402.4859}{{\ttfamily 1402.4859}}].

\bibitem{Ma:2025rnv}
S.~Ma, L.~Lehner, H.~Yang, L.~E. Kidder, H.~P. Pfeiffer and M.~A. Scheel,
  \emph{{Emergent Turbulence in Nonlinear Gravity}},
  \href{https://doi.org/10.1103/c9m4-mj3t}{\emph{Phys. Rev. Lett.} {\bfseries
  136} (2026) 061401} [\href{https://arxiv.org/abs/2508.13294}{{\ttfamily
  2508.13294}}].

\bibitem{Cardoso:2004nk}
V.~Cardoso, O.~J.~C. Dias, J.~P.~S. Lemos and S.~Yoshida, \emph{{The Black hole
  bomb and superradiant instabilities}},
  \href{https://doi.org/10.1103/PhysRevD.70.049903}{\emph{Phys. Rev. D}
  {\bfseries 70} (2004) 044039}
  [\href{https://arxiv.org/abs/hep-th/0404096}{{\ttfamily hep-th/0404096}}].

\bibitem{Loutrel:2020wbw}
N.~Loutrel, J.~L. Ripley, E.~Giorgi and F.~Pretorius, \emph{{Second Order
  Perturbations of Kerr Black Holes: Reconstruction of the Metric}},
  \href{https://doi.org/10.1103/PhysRevD.103.104017}{\emph{Phys. Rev. D}
  {\bfseries 103} (2021) 104017}
  [\href{https://arxiv.org/abs/2008.11770}{{\ttfamily 2008.11770}}].

\bibitem{Ripley:2020xby}
J.~L. Ripley, N.~Loutrel, E.~Giorgi and F.~Pretorius, \emph{{Numerical
  computation of second order vacuum perturbations of Kerr black holes}},
  \href{https://doi.org/10.1103/PhysRevD.103.104018}{\emph{Phys. Rev. D}
  {\bfseries 103} (2021) 104018}
  [\href{https://arxiv.org/abs/2010.00162}{{\ttfamily 2010.00162}}].

\end{thebibliography}\endgroup
\end{document}